\documentclass[lettersize,journal]{IEEEtran}
\usepackage{amsmath,epsfig,lipsum,booktabs,xcolor,mathtools}
\usepackage{hyperref}
\usepackage{amsfonts}
\usepackage{algorithmic}
\usepackage{algorithm}
\usepackage{epsfig}
\usepackage{times}
\usepackage[caption=false,font=normalsize,labelfont=sf,textfont=sf]{subfig}
\usepackage{textcomp}
\usepackage{stfloats}
\usepackage{url}
\usepackage{verbatim}
\usepackage{graphicx}
\usepackage{cite}
\usepackage{xcolor}
\usepackage{multicol}
\usepackage{cuted}
\usepackage{orcidlink}
\usepackage{tikz}
\newcommand{\rev}[1]{{\color{black}#1}}
\usepackage{balance}

\def\BibTeX{{\rm B\kern-.05em{\sc i\kern-.025em b}\kern-.08em
    T\kern-.1667em\lower.7ex\hbox{E}\kern-.125emX}}
\begin{document}

\title{Omnidirectional Video Super-Resolution using Deep Learning}
\author{Arbind Agrahari Baniya\orcidlink{0000-0002-9359-6506},  
Tsz-Kwan Lee\orcidlink{0000-0003-4176-2215}, Peter W. Eklund\orcidlink{0000-0003-2313-8603}, and Sunil Aryal\orcidlink{0000-0002-6639-6824}}

\maketitle

\begin{abstract}
  Omnidirectional Videos (or {360\textdegree} videos) are widely used in Virtual Reality (VR) to facilitate immersive and interactive viewing experiences. However, the limited spatial resolution in {360\textdegree} videos does not allow for each degree of view to be represented with adequate pixels, limiting the visual quality offered in the immersive experience. Deep learning Video Super-Resolution (VSR) techniques used for conventional videos could provide a promising software-based solution; however, these techniques do not tackle the distortion present in equirectangular projections of {360\textdegree} video signals. An additional obstacle is the limited {360\textdegree} video datasets to study. To address these issues, this paper creates a novel 360\textdegree~Video Dataset (360VDS) with a study of the extensibility of conventional VSR models to {360\textdegree} videos. This paper further proposes a novel deep learning model for {360\textdegree} Video Super-Resolution ({360\textdegree} VSR), called \rev{Spherical Signal Super-resolution with a Proportioned Optimisation (S3PO)}. S3PO adopts recurrent modelling with an attention mechanism, unbound from conventional VSR techniques like alignment. With a purpose-built feature extractor and a novel loss-function addressing spherical distortion, S3PO outperforms most state-of-the-art conventional VSR models and 360\textdegree~specific super-resolution models on {360\textdegree} video datasets. A step-wise ablation study is presented to understand and demonstrate the impact of the chosen architectural sub-components, targeted training and optimisation.

\end{abstract}


\section{Introduction}
\begin{figure}[ht!]
\vspace{-1em}
         \centering
         \includegraphics[width = 0.46 \textwidth]{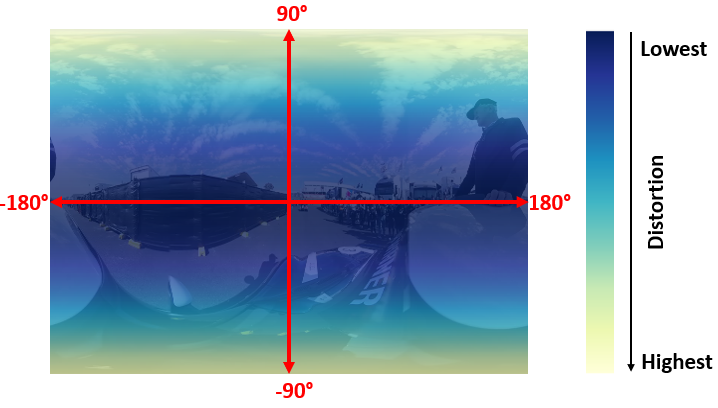} 
         \vspace{-1em}
         \caption{Illustration of an EquiRectangular Projection (ERP) frame with corresponding wide Field-of-View (FoV) and the induced distortion map.}
         \label{fig:erp}
         \vspace{-0.8em}
    \end{figure}

\IEEEPARstart{3}{60\textdegree} videos are increasingly popular, rapidly becoming the preferred format for multimedia in Virtual Reality (VR)~\cite{maniotis2019tile,yaqoob2020survey}.
Also called omnidirectional videos, spherical videos or panoramic videos; {360\textdegree} videos consist of {360\textdegree} of horizontal and {180\textdegree} of vertical Field of View (FoV). 
{360\textdegree} videos are primarily used for creating immersive experiences for their viewers by allowing up to six degrees of freedom of movement when interacting within the virtual environment. These videos are created using either a single camera with multiple sensors or multiple single-sensor cameras. The views captured from each sensor are stitched to create a single omnidirectional view. The spherical signal is then projected on a rectangular plane, by mapping the yaw and pitch, producing \rev{an EquiRectangular Projection (ERP)}. While other forms of projection for {360\textdegree} signals exist, such as cube map projection, equirectangular is the most widely used projection~\cite{zhou2019video} and is the format studied in this work. An example of the EquiRectangular Projection (ERP) frame is illustrated in Fig.~\ref{fig:erp} which shows the wide FoV and distortion present in ERPs resulting from mapping the spherical signal on a rectangular plane.

For a similar viewing experience to 1080p High-Definition (HD) conventional video, a resolution of $3840\times 2160$ ($= [1920\times1080]\times 4$) pixels is recommended by YouTube VR for {360\textdegree} videos~\cite{10.1145/3097895.3097896}. Similarly, in light of the wider view plane, {360\textdegree} videos require $\times8$ more data to be transmitted than conventional videos for a similar level of perceptual quality~\cite{9155477}. To mimic human biological perception, 60 pixels are needed to represent each degree of view~\cite{8329628}. \rev{This means for true immersive impact, {360\textdegree} of horizontal FoV needs to be represented by 21,600 pixels.} Thus the key factor inhibiting the adoption of {360\textdegree} videos in immersive contexts is the spatial resolution of the format. To bridge this gap, we explore the idea of enhancing {360\textdegree} videos using deep learning-based video super-resolution (VSR) technology. 

Recent advances in VSR for conventional videos show competent results, specifically when enhancing low-resolution videos by up to $\times4$ higher resolutions~\cite{RBPN,EDVR,RSDN,BasicVSR}. Such a software-based solution, built specifically for {360\textdegree} video context, could address the spatial resolution-related limitations in the {360\textdegree} video domain. Therefore, in this work, we explore the super-resolution of {360\textdegree} videos to achieve a $\times4$ spatial resolution enhancement of EquiRectangular Projection (ERP) of these videos. To experiment with these issues, we create a novel {360\textdegree} video dataset to measure model performance against the super-resolution task. The {360\textdegree} Video Dataset (360VDS) introduced in this paper presents a large collection of 590 \rev{video clips in ERP format} with a varied range of Spatial Index (SI) and Temporal Index (TI), as discussed in Sec.~\ref{sec:data} ({360\textdegree} Video Dataset). A new deep learning-based {360\textdegree} Video Super-Resolution ({360\textdegree} VSR) model called \rev{Spherical Signal Super-resolution with a Proportioned Optimisation (S3PO)} is then proposed to model the task of super-resolution specifically for panoramic videos, as discussed in Sec.~\ref{sec:model} (Proposed {360\textdegree} Video Super-Resolution Model). To address the limitations of existing VSR models, targeted recurrent modelling with a {360\textdegree} Feature Extractor and optimisation using a novel loss function are also introduced. 

Our empirical evaluation and analysis show that although conventional VSR models perform satisfactorily on {360\textdegree} videos, they can be further improved with targeted modelling and training. Our proposed S3PO model outperforms most of the existing state-of-the-art conventional and {360\textdegree} VSR models in quality evaluation matrices specific to both conventional and {360\textdegree} signals. \rev{ Furthermore, a detailed ablation study presented in Sec.~\ref{sec:ab_study} (Ablation Study) demonstrates comparative benefits of the strategic architectural choices, training with domain adaptation and optimisation of the S3PO model with the novel loss function.} 

The \rev{five-fold} contributions in this paper are summarised as:
\begin{enumerate}
    \item development of a new {360\textdegree} video dataset with more diverse spatial and temporal contexts than the existing ones for benchmarking {360\textdegree} VSR algorithms.
    \item study of the extensibility of existing conventional VSR models for {360\textdegree} videos, discussed in Sec.~\ref{sec:quant_eval} (Quantitative Evaluation).
    \item introduction of a novel deep learning-based {360\textdegree} VSR model, called S3PO, with: 
        \begin{enumerate}
            \item a hybrid recurrent architecture incorporating a sliding window and dual-duct residual blocks for effectively utilizing both local and global information;
            \item \rev{a panorama-specific custom-designed feature extractor with an attention mechanism for local feature extraction and information replenishment; }
            \item domain adaptation of super-resolution task from conventional videos to {360\textdegree} videos;
            \item weighted spherically smooth L1 loss function for distortion-aware super-resolution.
        \end{enumerate}
    \item state-of-the-art performance in super-resolution of {360\textdegree} videos on existing and new {360\textdegree} video datasets, discussed in Sec.~\ref{sec:quant_eval}(Quantitative Evaluation) and ~\ref{sec:qual_eval}(Visual Comparison).
    \item \rev{a detailed ablation study investigating the impact of architectural and learning choices for {360\textdegree} VSR along with the influences of conventional vs. {360\textdegree}-specific methodologies.}
\end{enumerate}

\section{Background}
      \subsection{Conventional Video Super-Resolution}
 Deep learning has been widely adopted to show how high-resolution (HR) output can be generated with improved  quality from low-resolution (LR) conventional video inputs~\cite{EDVR,TGA,RSDN}. The learning-based methods for conventional VSR typically consist of four key components, namely feature extraction, alignment, and fusion, followed by reconstruction and up-sampling. Among these, extracting relevant features from accurately aligned frames and fusing them are tasks of key significance for learning spatiotemporal correlation in a given temporal radius~\cite{RBPN,BasicVSR}. Techniques such as Motion Estimation and Motion Compensation (MEMC), using optical flow followed by warping~\cite{RBPN} or deformable convolution~\cite{deformConv}, are commonly used for either explicit (or implicit) alignment, respectively. Alternative to these, only 2D/3D convolutions~\cite{DUF,FeatGAN} and Recurrent Neural Networks (RNNs) have been used to extract features from unaligned frames and learn the sequential nature of video in doing so.

     \begin{figure}[ht!]
         \centering
         \includegraphics[width = 0.44\textwidth]{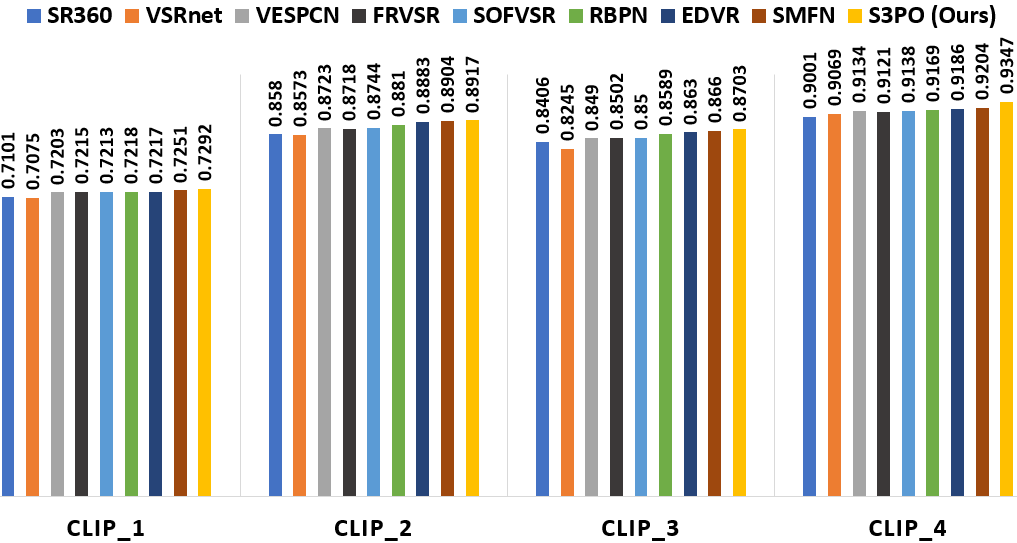} 
         \caption{WS-SSIM results on four clips of MiG Panorama Testset~\cite{liu2020single}.}
         \label{Fig:ssim_chart}
    \end{figure}
\begin{figure}[ht!]
         \centering
         \includegraphics[width = 0.44\textwidth]{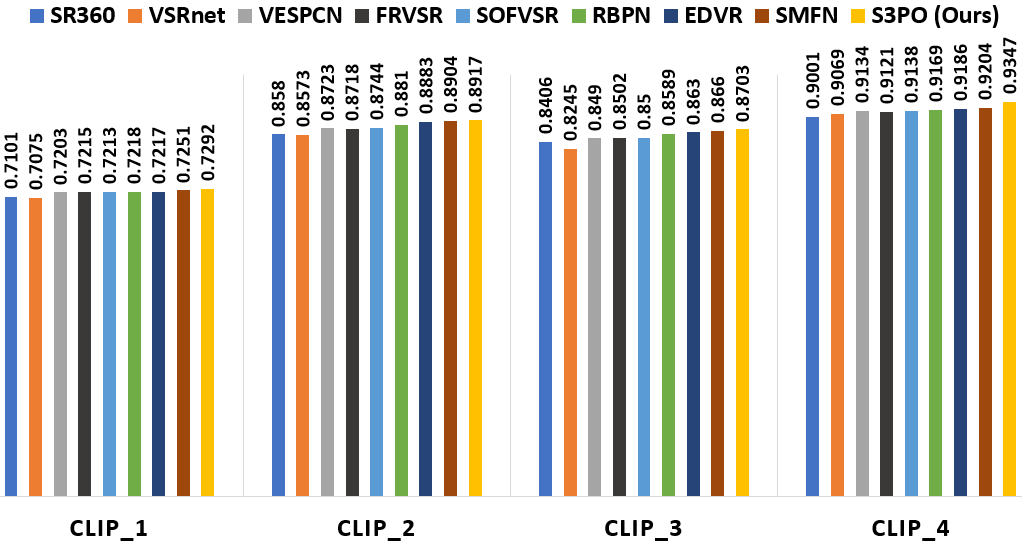}  
          \vspace{-0.7em}
         \caption{SSIM results on four clips of MiG Panorama Testset~\cite{liu2020single}.}
         \label{Fig:wsssim_chart}
    \end{figure}
    
   Recurrent Neural Networks (RNNs) are gaining popularity in the conventional Video Super Resolution (VSR) space thanks to their proven ability in effective sequential modelling. Recent RNN-based models, such as Recurrent Residual Network (RRN)~\cite{RRN}, Recurrent Structure-Detail Network (RSDN)~\cite{RSDN}, \rev{ BasicVSR~\cite{BasicVSR}, Replenished Recurrency with Dual-Duct (R2D2)~\cite{R2D2} and BasicVSR++~\cite{basic++}}, have demonstrated the capacity to learn long-term inter-frame correlation within a given video and consequently improve super-resolution quality. The global information propagation in RNNs offers a time and cost-efficient alternative to their alignment-based counterparts.
   
    \rev{There is an increasing trend of leveraging global information present in videos using bidirectional recurrent models~\cite{BasicVSR,basic++,yi2021omniscient}. These models make use of an all-frames-in approach to super-resolve the input videos. Although the all-frames-in approach has shown promising results~\cite{BasicVSR,basic++}, it can be computationally expensive due to the concurrent processing of information bidirectionally (forward and backward) across a long temporal radius for a video~\cite{R2D2}. Among the bidirectional recurrent model, the order of information propagation is increasingly becoming complex with not just two flows of information (forward and backward) but multiple layers of these flows and their crossovers, as seen in BasicVSR++~\cite{basic++}. This limits the applicability of such models to offline settings only, as in most online settings, all frames are not available at a given time. 
    
    More sophisticated usage of hybrid structures incorporating unidirectional recurrent models with replenishing information and multi-staged refinement is emerging to mitigate issues like vanishing gradient from unidirectional recurrent networks~\cite{R2D2}. These models are showing promising results when compared to the bidirectional counterparts while still being applicable in both online and offline application settings.}
    
    However, the extensibility of any recurrent modelling, and its effectiveness for {360\textdegree} VSR, has yet to be studied. In this work, we aim to address this gap by firstly testing state-of-the-art conventional VSR models on our newly created 360VDS dataset. We prove that a purpose-built {360\textdegree} VSR model can outperform conventional VSR technologies, which are not well-suited to the format because of the large FoV and the presence of distortion and \rev{horizontal cyclicity} in EquiRectangular Projections (ERPs).

    \begin{figure*}[ht!]
    \vspace{-1em}
         \centering
         \includegraphics[width = 0.70\textwidth]{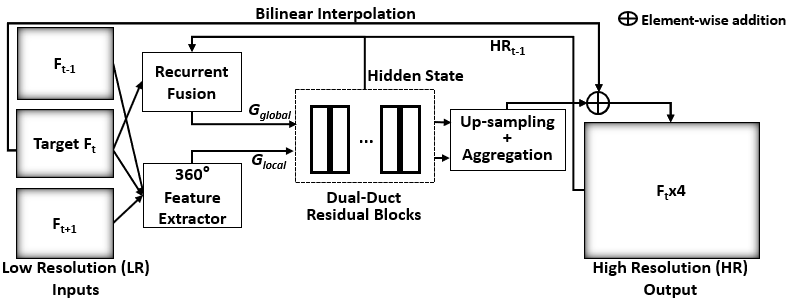}
          \vspace{-0.8em}
         \caption{Block diagram of the key architecture of the proposed \rev{Spherical Signal Super-resolution with a Proportioned Optimisation (S3PO) model}.}
         \label{Fig:model}
         \vspace{-1em}
    \end{figure*}

    \subsection{{360\textdegree} Super-Resolution}
    \label{sec:360_background}
  
    Super-resolution technologies have been applied for  omnidirectional images by pioneers like Fakour-Sevom {\it et al.}~\cite{fakour2018360}, Ozcinar {\it et al.}~\cite{8901764} and Nishiyam {\it et al.}~\cite{9506233}. These studies reveal that Single Image Super-Resolution (SISR) can be applied to panoramic images. In these approaches, existing SISR models for conventional images~\cite{dong2015image} have been fine-tuned by using either {360\textdegree} image datasets~\cite{xiao2012recognizing}, re-training to optimise distortion-aware loss functions~\cite{8901764}, or by applying distortion maps on input images before being fed into the network~\cite{9506233}.
    
    However, there is limited work extending super-resolution to {360\textdegree} video signals. Dasari {\it et al.}~\cite{dasari2020streaming} proposed adopting VSR in {360\textdegree} videos to mitigate the \rev{bandwidth-related} requirements for an adaptive video streaming service. A micro-model for super-resolution was designed as part of a streaming system to enhance the spatial quality of compressed tiles by passing each of them through multiple convolution layers, followed by final deconvolution and upsampling. The Dasari {\it et al.} approach is based on bandwidth-related requirements for a streaming service; the video-specific properties were not considered in the super-resolution task where the temporal correlation between video frames is a crucial factor to boost the visual quality for higher resolution output.

    Liu {\it et al.}~\cite{liu2020single} focus on creating a dual network VSR model called Single and Multi-Frame Recurrent Network (SMFN). One pipeline is used for Single Image Super-Resolution (SISR), and the other for Multi-Image Super-Resolution (MISR). Approaching the {360\textdegree} VSR problem by combining SISR and MISR limits SMFN in learning spatiotemporal properties in a local neighbourhood of limited temporal radius. SMFN has been shown to generate better super-resolution outcomes for {360\textdegree} videos compared to conventional state-of-the-art VSR models, such as EDVR~\cite{EDVR} and RBPN~\cite{RBPN}. However, the MiG Panorama dataset, used in the study by Liu {\it et al.}~\cite{liu2020single},
    has only four clips in the test set, and most of the recent \rev{recurrent network-based} VSR models have not been used in their comparison; although, these models have proven to outperform EDVR and RBPN on several benchmark datasets due to their better sequential modelling ability. Fig.~\ref{Fig:ssim_chart} and Fig.~\ref{Fig:wsssim_chart} give a glimpse of S3PO's performance on the MiG Panorama test set with four clips. As shown in these figures, S3PO outperforms most of the existing VSR models in both Structural Similarity Index Measure (SSIM)~\cite{hore2010image} and Weighted Spherically-SSIM (WS-SSIM)~\cite{zhou2018weighted} metrics in each of the four clips of MiG Panorama test set. 
    
\section{Proposed {360\textdegree} Video Super-Resolution Model}
\label{sec:model}

    \subsection{Architectural Overview}

    Fig.~\ref{Fig:model} shows the architectural overview of our proposed \rev{Spherical Signal Super-resolution with a Proportioned Optimisation (S3PO)} model. S3PO adopts recurrent modelling in combination with a sliding-window mechanism. The recurrence allows for \rev{sequential modelling} by enabling global memory propagation across equirectangular video frames over time. For uninterrupted global memory propagation, the current target frame $F_t$ is directly fused with the hidden state $h_{t-1}$, and the super-resolved output $\mbox{\it HR}_{t-1}$ from the previous timestamp \rev{establishing a recurrent pathway}. \rev{Such recurrent neural networks have a proven ability to model the sequential nature of continuous spatiotemporal data such as videos. However, the propagation of global memory diminishes over time in unidirectional recurrent models, limiting their temporal receptivity and long-term sequential modelling ability. Additionally, any sudden change or noise within a given time stamp can disrupt the flow of global memory and introduce undesirable memory features. To mitigate these issues, an information replenishment mechanism with a sliding window is used in the S3PO model. This} allows highly correlated local features to be extracted and used for information replenishment in the recurrent network. To capture local features \rev{from a given sliding window}, we introduce a novel {360\textdegree} Feature Extractor specifically designed to extract joint features from three consecutive panoramic frames, $F_{t-1}$, $F_t$ and $F_{t+1}$.

    The S3PO model aims to replace the \rev{computationally intensive and error-prone} conventional alignment steps with the sequential modelling ability of a recurrent architecture, as illustrated in Fig.~\ref{Fig:model}. The {360\textdegree} Feature Extractor with attention mechanism replenishes vanishing memory with highly correlated features extracted locally. \rev{The co-joint feature extraction in the proposed feature extractor, followed by the attention mechanism, allows S3PO to extract relevant features directly from unaligned frames without the need for explicit alignment and further to pay varied attention to these extracted features across both spatial and temporal dimensions.} The extracted and fused features from both local and global propagation, as denoted in Fig.~\ref{Fig:model} respectively as $G_{local}$ and $G_{global}$, are then propagated through a network of dual-duct residual blocks. Each dual-duct residual block refines the features individually while still making use of mutual information exchange to allow both local and global memory to be refined with respect to each other. \rev{Such a network of dual-ducts allows different frequency details to be captured separately while still fostering meaningful feature representations mutually}. The two sets of features obtained from the last residual block are subjected to pixel-shuffle operations to enable depth-to-space transformation of the obtained {360\textdegree} feature maps. The resultant $\times4$ feature map is then used as residue, added element-wise to a bi-linearly interpolated target frame to produce the final $\times 4$ super-resolved equirectangular frame $\mbox{HR}_t$.

    \begin{figure*}[t]
     \centering
     \includegraphics[width = 0.97\textwidth]{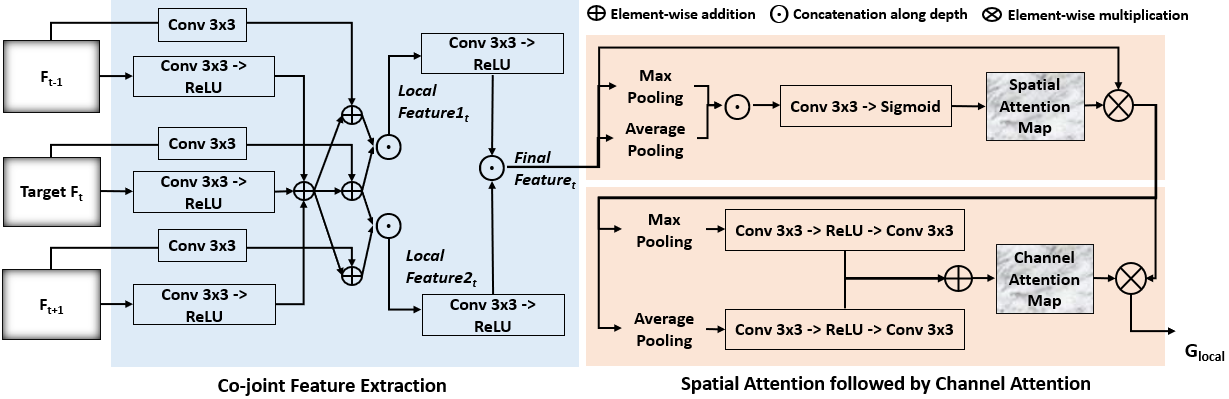}
     \vspace{-0.5em}
     \caption{{360\textdegree} Feature Extractor with co-joint feature extraction (see Eqn.~(\ref{eqn:feat_extract1})) and self-attention mechanism (see Eqn.~(\ref{eqn:feat_extract2})) }.
     \label{Fig:360FeatExtract}
     \vspace{-1.5em}
\end{figure*} 
 \begin{figure*}[hb]
        
    \begin{equation}
    \label{eqn:feat_extract1}
    \begin{gathered}
    {\mbox{\it Joint\;Feature}_{t}} =\mbox{ReLU}(\mbox{\it Conv}_{3\times3}(F_{t-1})) \oplus \mbox{ReLU}(\mbox{\it Conv}_{3\times3}(F_{t})) \oplus \mbox{ReLU}(\mbox{\it Conv}_{3\times3}(F_{t+1})) \\
    {\mbox{\it Local\;Feature1}_{t}} = {\mbox{\it  Correlated\;Feature}} (F_{t-1}) \odot {\mbox{\it  Correlated\;Feature}} (F_{t}) ~ {\mbox and}\\
    {\mbox{\it Local\;Feature2}_{t}} = {\mbox{\it  Correlated\;Feature}} (F_{t+1}) \odot {\mbox{\it  Correlated\;Feature}} (F_{t}) \\
    \text{where, } {\mbox{\it Correlated\;Feature}} ( \cdot) = {\mbox{\it Joint\;Feature}_{t}} \oplus \mbox{\it Conv}_{3\times3}(\cdot)\\
    {\mbox{\it Final\;Feature}_{t}} = \mbox{ReLU}(\mbox{\it Conv}_{3\times3}({\mbox{\it Local\;Feature1}_{t}} )) \oplus \mbox{ReLU}(\mbox{\it Conv}_{3\times3}({\mbox{\it Local\;Feature2}_{t}}))
    \end{gathered}
    \end{equation}
        \end{figure*}

     \begin{figure*}[ht]
     \vspace{-1em}
    \begin{equation}
    \label{eqn:feat_extract2}
    \begin{gathered}
     {\mbox{\it Spatial-Attention\;Map}_{t}} = \sigma (\mbox{\it Conv}_{3\times3}({\mbox{\it Max\;Pool}}({\mbox{\it Final\;Feature}_{t}}) \odot {\mbox{\it Avg\;Pool}}({\mbox{\it Final\;Feature}_{t}})))\\
     {\mbox{\it Spatial-Attention\;Feature}_{t}} = {\mbox{\it Spatial-Attention\;Map}_{t}} \otimes {\mbox{\it Final\;Feature}_{t}}\\
     {\mbox{\it Channel-Attention\;Map}_{t}} = {\mbox{\it CA\;Feat}}({\mbox{\it Max\;Pool}}({\mbox{\it Spatial-Attention\;Feature}_{t}})) \oplus {\mbox{\it CA\;Feat}}({\mbox{\it Avg\;Pool}}({\mbox{\it Spatial-Attention\;Feature}_{t}}))\\
     \text{where, }{\mbox{\it CA\;Feat}}(\cdot) = \mbox{\it Conv}_{3\times3}(\mbox{ReLU} (\mbox{\it Conv}_{3\times3}(\cdot))\\
     G_{\mbox{\it local}} = {\mbox{\it Channel-Attention\;Map}_{t}} \otimes {\mbox{\it Spatial-Attention\;Feature}_{t}}
    \end{gathered}
    \end{equation}
    \vspace{-3em}
    \end{figure*}

\subsection{{360\textdegree} Feature Extractor}
\label{sec:feat_extractor}

Alignment methods, such as MEMC, designed for conventional videos, do not consider the distortion present in EquiRectangular Projection (ERP)  frames. Optical Flow estimation~\cite{spynet}, primarily used for motion estimation, is an error-prone process, even for conventional videos when there are large motion or intensity changes between consecutive frames. With the added \rev{complexity} of distorted pixels, conventional methods can not be directly applied for accurate flow estimation in the context of {360\textdegree} videos. The objects in ERP frames have different levels of distortion at different latitudes. So, when an object moves between consecutive frames, distortion across different regions within the same object may also vary, making flow estimation between consecutive ERP frames challenging. Additionally, spherical boundaries of {360\textdegree} videos allow for cyclic motion. Although the geometric meaning of displacement is independent of the motion trajectory, cyclic motion means multiple possible paths between origin and target points in a rectangular plane, causing misleading or unsatisfactory estimation of flow using conventional methods~\cite{10.1145/3411764.3445499,bhandari2021revisiting,shi2022panoflow}.

 Rather than using conventional alignment~\cite{spynet}, we propose using learning-based technologies to directly extract \rev{relevant features from unaligned consecutive frames} in {360\textdegree} videos. Direct extraction and fusion of 2D features from video frames have proven to result in subpar performance when compared to features extracted from aligned frames in conventional videos~\cite{lucas2019generative}. The proposed {360\textdegree} Feature Extractor addresses this limitation in the specific context of {360\textdegree} videos using \rev{a two-staged feature extraction mechanism}.

As shown in Fig.~\ref{Fig:360FeatExtract}, the initial 2D features directly extracted from each EquiRectangular Projection (ERP) frame are refined with respect to \rev{joint features extracted from the combination of all three frames} in the local neighbourhood. This enables the extraction of mutually correlated local features, represented by $\mbox{\it Correlated\;Feature}$ in eqn.~(\ref{eqn:feat_extract1}). 
\rev{For each pair of the neighbouring frames ($F_{t-1}$ and $F_{t+1}$), corresponding features are then further refined with respect to the target frame ($F_t$) features}. This allows for meaningful feature extraction from neighbouring frames that correlate with the target frame. The final two sets of features extracted are then fused to generate a single set of local features represented by $\mbox{\it Final Feature}_{t}$ in Eqn.~(\ref{eqn:feat_extract1}). Each step involved in the first stage of feature extraction is outlined in  Eqn.~(\ref{eqn:feat_extract1}), where $\mbox{\it Conv}_{3\times3}$ is the convolution operation with filter-size $3\times3$, `$\odot$' represents the concatenation operation along the feature depth, `$\oplus$' represents the element-wise addition operation and ReLU represents the Rectified Linear Unit. The single set of $\mbox{\it Final Feature}_{t}$, obtained from the first stage for a given timestamp $t$, is then propagated to the second stage of the {360\textdegree} Feature Extractor.

In the second phase, to cater for the varied spatial differences caused by distortions present in ERP frames, we propose to use a spatial attention mechanism~\cite{woo2018cbam} to allow the S3PO model to assign varied attention to different spatial regions within the distorted ERP frames. Additionally, a channel attention mechanism~\cite{woo2018cbam} is also used to allow the model to differentiate between the features extracted from the temporally co-located frames. The attention mechanism involved in the second stage of feature extraction is outlined in Eqn.~(\ref{eqn:feat_extract2}), where `$\otimes$' represents the element-wise multiplication operation and $\sigma $ represents the sigmoid activation function. This stage allows {360\textdegree} Feature Extractor to learn to extract spatiotemporally correlated features representing the unique nature of {360\textdegree} video frames while extracting mutually correlated features in a given local temporal radius. \rev{By combining and correlating the features from temporally co-located frames, the {360\textdegree} feature extractor enables a comprehensive representation of the information in the given temporal radius, including motion information. This allows S3PO to be free from \rev{conventional} alignment steps, which are known to be error-prone and computationally intensive in the context of conventional videos and thus do not extend well for the unique nature of {360\textdegree} videos~\cite{bhandari2021revisiting}}. A single set of extracted features $G_{\mbox{\it local}}$, for the given timestamp $t$, is then propagated for further refinement with respect to the global memory.
\begin{figure}[ht!]
     \centering
     \includegraphics[width = 0.438\textwidth]{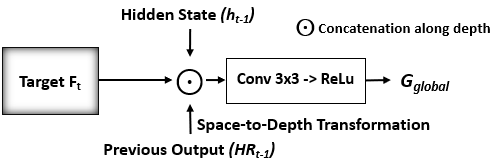}   
      \vspace{-0.8em}
     \caption{Recurrent global information fusion in the S3PO model (see Eqn.~(\ref{eq:1})).}
     \label{Fig:global_fusion}
     \vspace{-1em}
\end{figure}

    \subsection{Recurrent Global Fusion}

    The S3PO model learns the long-term temporal reliance between ERP frames by propagating global memory. For this, it uses hidden-state information $h_{t-1}$ and the super-resolved output from previous timestamp $\mbox{HR}_{t-1}$, as illustrated in Fig.~\ref{Fig:global_fusion}. The current target frame $F_t$ is directly fused with the hidden-state, and $\times4$ space-to-depth transformed (using a pixel-unshuffle operation) version of $\mbox{HR}_{t-1}$, to obtain a single set of fused features. This can be formally denoted by Eqn.~(\ref{eq:1}), where $G_{\mbox{\it global}}$ is the finally fused features representing the global information flow in recurrent architecture, and $\downarrow$ represents a pixel-unshuffle operation. The grouping of hidden state and output from the previous timestamp with the current LR target input allows the S3PO model to learn temporal correlations between frames across the video and allows long-term propagation of texture and context details, enabling the sequential modelling desired from an RNN for VSR.
    
    \begin{equation}\label{eq:1}
     G_{\mbox{\it global}} = \mbox{ReLU}(\mbox{\it Conv}_{3\times3}[F_t \odot h_{t-1} \odot (\mbox{HR}_{t-1})_{\downarrow}])
    \end{equation}

    \subsection{Dual-Duct Refinement}
    
    The two sets of features $G_{\mbox{\it local}}$ and $G_{\mbox{\it global}}$ are the data propagation results respectively depicted in Fig.~\ref{Fig:360FeatExtract} and Fig.~\ref{Fig:global_fusion}. They are forwarded for refinement through a residual convolution network. This network consists of ten residual blocks, where each block consists of a dual duct to learn global and local relationships between ERP frames concurrently. 
    \vspace{0.8em}
    \begin{figure}[ht!]
        \vspace{-1em}
         \centering
         \includegraphics[width = 0.405\textwidth]{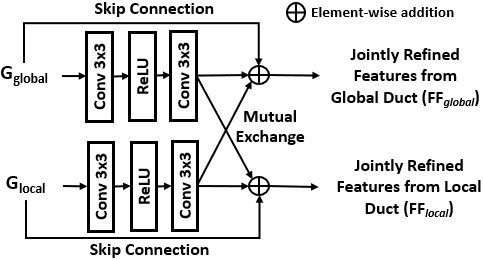}
         \vspace{-0.8em}
         \caption{Reconstruction and refinement using dual-duct with mutual information exchange (see eqns.~(\ref{eq:6}-\ref{eq:5})).}
         \label{Fig:res_block}
         
    \end{figure}

    Rather than building two completely independent pipelines, we build residual blocks where both the local and global ducts are leveraged via mutual information exchange as shown in Fig.~\ref{Fig:res_block} and eqn.~(\ref{eq:6}). Each duct consists of a $3\times3$ convolution operation followed by a ReLU activation and additionally a $3\times3$ convolution operation as shown in Fig.~\ref{Fig:res_block}. The features obtained from each duct are added element-wise with the respective identity feature obtained through the skip connection. Skip connections of features, in conjunction with the jointly learned refinement, allows for the construction of a deeper convolution network while enabling the network to be robust to vanishing gradient, effectively propagating both local and global information while mitigating error accumulation.
    
    The two final sets of jointly refined features from the last residual blocks $\mbox{\it FF}_{\mbox{\it local}}$ and $\mbox{\it FF}_{\mbox{\it global}}$ are generated as shown in Fig.~\ref{Fig:res_block}. These two sets of features are then used to create hidden-state information $h_t$ for the given time-stamp $t$ to be used for the future time-stamp $t+1$, as defined in eqn.~(\ref{eq:4}). The two finally refined features $LF_t$ and $GF_t$, \rev{from the local and global information propagation} respectively, are obtained from $\mbox{\it FF}_{\mbox{\it local}}$ and $\mbox{\it FF}_{\mbox{\it global}}$ respectively (as shown in eqns.~(\ref{eq:6}-\ref{eq:5})).
    
    \begin{equation}\label{eq:6}
    \begin{split}
    & \mbox{\it FF}_{\mbox{\it local}} = G_{\mbox{\it local}} \oplus R(G_{\mbox{\it local}})\oplus R( G_{\mbox{\it global}}) \\
    & \mbox{\it FF}_{\mbox{\it global}} = G_{\mbox{\it global}} \oplus R(G_{\mbox{\it global}})\oplus R( G_{\mbox{\it local}}) \\
    & \mbox{where, } R(\cdot) = \mbox{\it Conv}_{3\times3}(ReLU(\mbox{\it Conv}_{3\times3}(\cdot)))~ \mbox{and}\\
    \end{split}
    \end{equation}
    \begin{equation}\label{eq:4}
     h_t = \mbox{ReLU}(\mbox{\it Conv}_{3\times3}[\mbox{\it FF}_{\mbox{\it local}}]) \oplus \mbox{ReLU}(\mbox{\it Conv}_{3\times3}[\mbox{\it FF}_{\mbox{\it global}}])
    \end{equation}

    \begin{equation}\label{eq:5}
    \mbox{\it LF}_t = \mbox{\it Conv}_{3\times3}[ \mbox{\it FF}_{\mbox{\it local}}],
     \mbox{\it GF}_t = \mbox{\it Conv}_{3\times3}[ \mbox{\it FF}_{\mbox{\it global}}]
     \vspace{-1em}
    \end{equation}

     \begin{figure}
     \centering
     \includegraphics[width=0.49\textwidth]{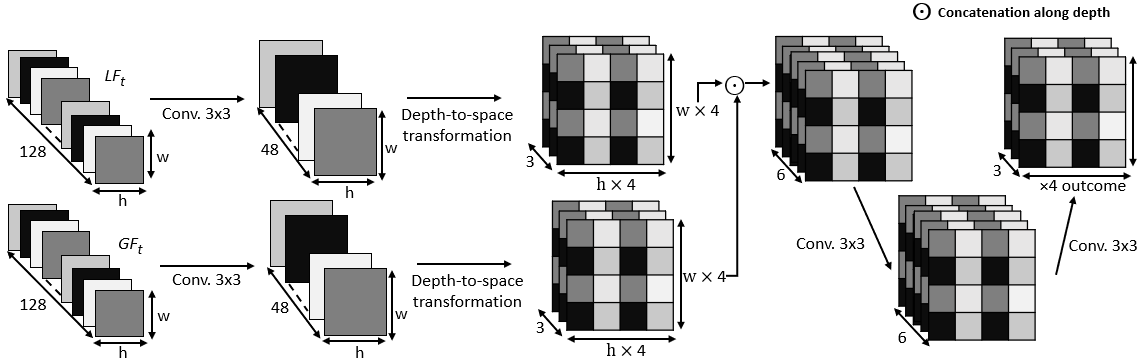}
      \vspace{-1.5em}
     \caption{Up-sampling and aggregation in S3PO model.}
     \label{Fig:pixel_shuffle}
      \vspace{-1em}
    \end{figure}

    \subsection{Upsampling}
    \begin{figure*}[ht]
         \centering
         \includegraphics[width = 0.96\textwidth]{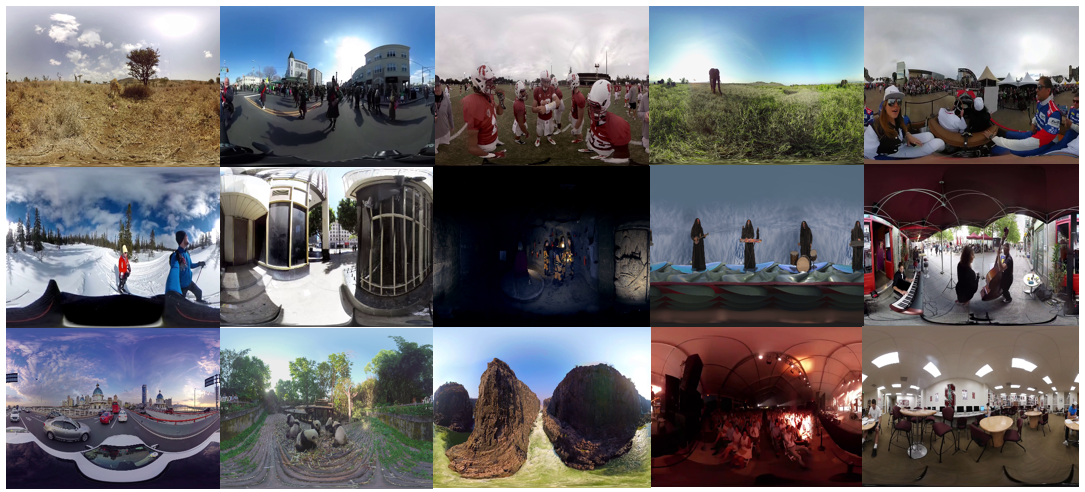}  
        \vspace{-0.8em}
         \caption{Randomly sampled EquiRectangular Project (ERP) frames with uniform resolution representing different video clips from the newly created {360\textdegree} Video Dataset (360VDS) showcasing diverse contents and lighting conditions.}
         \label{Fig:dataset}
    \end{figure*}
    Following the refinement operation in the residual blocks, the two outcomes from local and global-duct, namely $ \mbox{\it LF}_t$ and $ \mbox{\it GF}_t$ respectively, are then subjected to the $3\times3$ convolution operation followed by a depth-to-space transformation in order to cast the $\times4$ feature depth to spatial data as shown in Fig.~\ref{Fig:pixel_shuffle}. For depth-to-space transformation, the S3PO model uses the pixel-shuffle operation. The two outputs, each with three channel features then obtained, corresponding to the super-resolved residues from the two groups of information propagation. Finally, the two outputs are concatenated along depth and passed through two $3\times3$ convolution layers to obtain the $\times4$ super-resolved residue. This spatially super-resolved residue is then added element-wise with a bi-linearly interpolated LR frame for the final HR reconstruction \rev{($F_t \times 4$)} of the given $F_t$.

\section{{360\textdegree} Video Dataset}
\label{sec:data}
    \subsection{Data Collection}
    {360\textdegree} video is an emerging format in multimedia. There is, therefore, presently a lack of benchmark datasets that can be used for standard training and model evaluation. As discussed earlier in Sec.~\ref{sec:360_background} ({360\textdegree} Super-Resolution), there are only two known works that have attempted super-resolution for {360\textdegree} videos. And only one of those has created a dataset that could be used generally for {360\textdegree} video super-resolution research. However, the MiG Panorama Video dataset~\cite{liu2020single} in question  consists of  only $204$ {360\textdegree} video clips, and only four of these are publicly available and used for testing. This limits the diversity of motion, content and light conditions represented by the test set. 
    
    To address this, we create a new dataset of panoramic videos specifically designed for super-resolution called the {360\textdegree} video dataset (360VDS). Open-source datasets used in other areas of {360\textdegree} video research are assembled to create the 360VDS, such as those used in quality assessment~\cite{li2018bridge}, compression~\cite{su2018learning}, salience modeling~\cite{10.1145/3304109.3325820}, along with those in surveys and literature studies~\cite{chiariotti2021survey,xu2020state,10.1145/3304109.3325812}. Additionally, we also make use of the publicly available {360\textdegree} video dataset from the Stanford VR lab called psych-360~\cite{miller2020personal}.
    
    In collecting this diverse range of open-source {360\textdegree} videos, we ensure that the number of moving objects ranges from none to single to multiple objects. At the same time, we also ensure that the videos selected contain different camera motions -- either fixed or movement along a single axis -- or rotation -- or a combination of these movements. The contents captured are also diverse, ranging from animals, trees/grasses, buildings, humans, day-to-day objects and even synthetic content. With this collection, we ensure a diverse representation of content and contexts, variable lighting conditions and different types of motion. In total, 301 videos were collected with variable duration. 
    
    \subsection{Dataset Formation}
    \begin{figure}[ht]
         \centering
         \includegraphics[width = 0.45\textwidth]{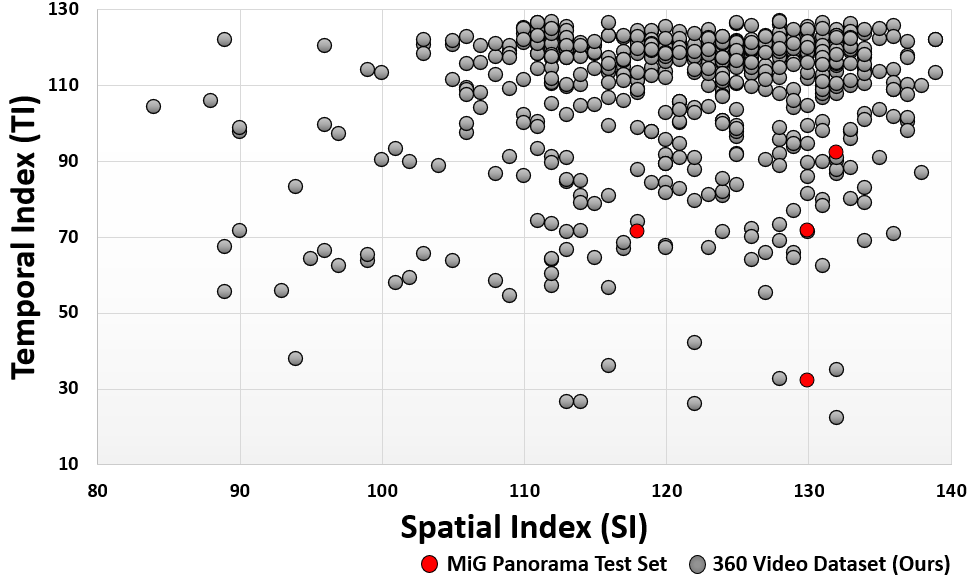}   
          \vspace{-0.8em}
         \caption{A representation of the spatial and temporal complexity present in the assembled 360VDS dataset.}
          \vspace{-0.8em}
         \label{Fig:si_ti}
    \end{figure}
     Each of the  videos in the  collection is then processed to detect scenes followed by segregation into multiple clips, each representing a single shot. {\sl PySceneDetect}~\cite{pyscene} was used to identify jump cuts in the collected videos. We use an adaptive content detector to convert RGB videos to HSV and compare the rolling average difference across all channels between adjacent frames. Shot detection, identified in this way, avoids false detection that would otherwise result from large camera motion.
     
     In total, 1,133 clips were created from the 301 collected videos. The 1,133 clips were then hand-crafted to remove any shots with no content, static content, or solely textual content. 896 video clips obtained in this way were then further filtered to remove shots with highly similar content and scenes. The result is 590 videos. However, most of these videos are too long and large, therefore, unsuitable for deep learning training and testing due to hardware demands. 
     
     To form the final benchmark dataset, up to 20 frames were extracted from each of the 590 videos. Additionally, all 590 videos were transformed to $480\times360$ pixels for \rev{consistent and uniform resolution}. Resizing also helps resolve the size-related bottleneck we would otherwise face when training our recurrent model. The 590 video dataset is then split randomly to select 45 clips as a test set and the remaining 545 as the training set. Fig.~\ref{Fig:dataset} shows some randomly sampled frames from the final dataset. The 360VDS dataset will be publicly available at  \href{https://github.com/arbind95/360VSR}{\color{blue}S3PO GitRepo} for the community to use as a benchmark dataset in future research and development. To ensure that the 360VDS represents diversity, the results of spatial and temporal complexity analysis, as per International Telecommunication Union's (ITU) standard guidelines~\cite{itu1999subjective}, is presented in Fig.~\ref{Fig:si_ti}. It is evident that 360VDS represents varied spatiotemporal complexity desired for training and evaluation in VSR domain~\cite{vimeo}. Furthermore, the four publicly available clips from the MiG Panorama test set are also analysed, as plotted in red in Fig.~\ref{Fig:si_ti}, to illustrate the need for a larger and more diverse test set.  

     \rev{Additionally, a test set with clips ranging from high-resolution to ultra-high-resolution is created to evaluate the performance of VSR models on a higher-resolution restoration task which poses more computational challenges. The REDS~\cite{REDS} test set and UDM10~\cite{UDM10} dataset, widely used for benchmarking in conventional video super-resolution, consist of four and ten HD clips, respectively. Our newly created 360 Ultra High-Definition (360UHD) dataset consists of eight clips ranging from HD to 4K; therefore, posing a similar benchmarking challenge to REDS and UDM10, however specifically designed for {360\textdegree} multimedia context. These clips are randomly selected from the $45$ clips of the 360VDS test set without being transformed to $480\times360$ pixels. Their original high-resolution ground truth is kept intact, ranging from HD to 4K, as presented in Table~\ref{tab:hq_result}. Testing the efficacy of omnidirectional video super-resolution (VSR) models on high-resolution videos is of utmost importance for the majority of {360\textdegree} multimedia applications, as the format requires a higher resolution to provide a truly seamless and immersive viewing experience. The process of super-resolving {360\textdegree} high-resolution videos, therefore, presents a greater challenge compared to conventional VSR since the models must preserve and enhance an increased amount of detail. Consequently, it is highly recommended that any {360\textdegree} VSR model be evaluated for super-resolving high-resolution videos. The details and evaluation results for each clip of the 360UHD test set is presented in Table~\ref{tab:hq_result}.}

    \begin{figure*}[ht]
    \vspace{-1.5em}
    \begin{equation}
    \label{eqn:loss}
    \mbox{\it WSS}-L1\;\mbox{\it Loss} =
        \begin{dcases}
            \sum_{i=1}^{\mbox{\it height}}\sum_{j=1}^{\mbox{\it width}} \left(\frac{0.5(\mbox{\it GT}_{i,j} - \mbox{\it HR}_{i,j})^2}{\beta}\right)\times \psi_{i,j}, & \text{if $| GT - HR |$}< \beta  \\
           \sum_{i=1}^{\mbox{\it height}}\sum_{j=1}^{\mbox{\it width}} ( |\mbox{\it GT}_{i,j}-\mbox{\it HR}_{i,j}|- 0.5\beta) \times \psi_{i,j} , & \text{otherwise}
        \end{dcases}
        \text{, where } \psi_{i,j} = \cos{\left({\frac{(i+0.5-height/2)\pi}{height}}\right)}
    \end{equation}
    \vspace{-2em}
    \end{figure*}
          \begin{figure}[ht]
         \centering
         \includegraphics[width = 0.488\textwidth]{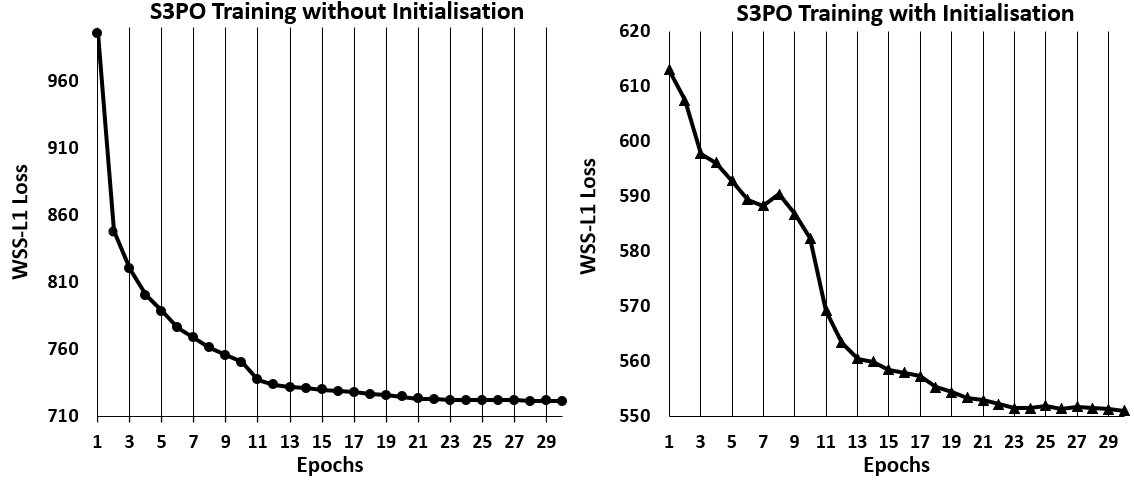} 
          \vspace{-2em}
         \caption{Comparison of WSS-L1 loss when the S3PO model is trained with (and without) being initialised with conventional VSR weights. }
         \label{Fig:training_curve}
         \vspace{-1em}
    \end{figure}
\section{Experiment and Evaluation}

    \subsection{Experimental Setting}

    \subsubsection{Domain Adaptation}
   The Vimeo90K~\cite{vimeo} dataset, a widely used benchmark dataset in conventional VSR, is used to train the S3PO model initially for a generic initialisation of the super-resolution task. The size and diversity presented by Vimeo90K allow the S3PO model to learn the task of super-resolution which is used as a knowledge base. This is then fine-tuned by retraining the S3PO model on 360VDS to allow for the specification of the generic initialisation in the context of {360\textdegree} videos. Out-of-domain knowledge has proven to be effective as a generic initialisation for new domains~\cite{thompson2018freezing}. Following the success of this approach in deep learning, we also make use of conventional VSR as a base initialisation to learn the task of {360\textdegree} video super-resolution more effectively. Fig.~\ref{Fig:training_curve} depicts the benefit of initialising the S3PO model with conventional VSR weights, this prevents early saturation of training and allows for further learning. The effectiveness of this approach is also demonstrated and discussed in Sec.~\ref{sec:ab_study} (Ablation Study).

    \subsubsection{Training}
    The input LR frames were generated using both Blur Degradation (BD) and Bicubic Interpolation (BI). For BD, we use Gaussian blur with a standard deviation of $\sigma = 1.6$ and $4\times$ down-sampling commonly used in the literature~\cite{liu2022video}. Using two forms of degradation allows the S3PO model to learn the super-resolution task for diverse scenarios and benchmarking against varied models.
    
    For the conventional VSR initialisation, weights were obtained by training the S3PO model on the conventional Vimeo90K~\cite{vimeo} dataset with a batch size of 8, for 70 epochs using Smooth-L1 loss~\cite{girshick2015fast}. Followed by this, in order to allow for variable length frames (up to 20 frames) in the 360VSD, a batch size of one is used for training the S3PO model on 360VDS during the 75 epochs at which point learning saturation occurs for both categories of inputs. S3PO makes use of the Adam optimizer~\cite{adam} to optimize the newly proposed Weighted Spherically Smooth-L1 loss, presented in Eqn.~(\ref{eqn:loss}) discussed in Sec.~\ref{sec:loss_function} for accurate pixel-to-pixel predictions specifically in the context of {360\textdegree} multimedia. The initial learning rate is set to $1 \times 10^{-4}$, and decayed by a factor of $10$ after every $10$ epoch. Model training and testing are performed using two NVIDIA Tesla V100 GPUs. The {\sl PyTorch}~\cite{NEURIPS2019_9015}-based source-code and training weights 
    will be made publicly available at \href{https://github.com/arbind95/360VSR}{\color{blue}S3PO GitRepo}.

    \subsubsection{Loss Function}
    \label{sec:loss_function}

    Conventional Smooth-L1 loss acts as both L1 and L2 losses conditioned to a hyper-parameter $\beta$. It combines the advantages of L1-loss (steady gradients for large values) and L2-loss (less oscillation during update when values are small). Thus, it is less sensitive to outliers and prevents exploding gradients in some cases. However, the loss functions used for conventional images and videos do not take into account the unique nature of EquiRectangular Projection (ERP)  frames. Distortion across the latitude present in ERP frames can cause learning-based models to be easily influenced by high errors in prediction across polar regions.

\begin{figure}[hb]
         \centering
         \includegraphics[width = 0.45\textwidth]{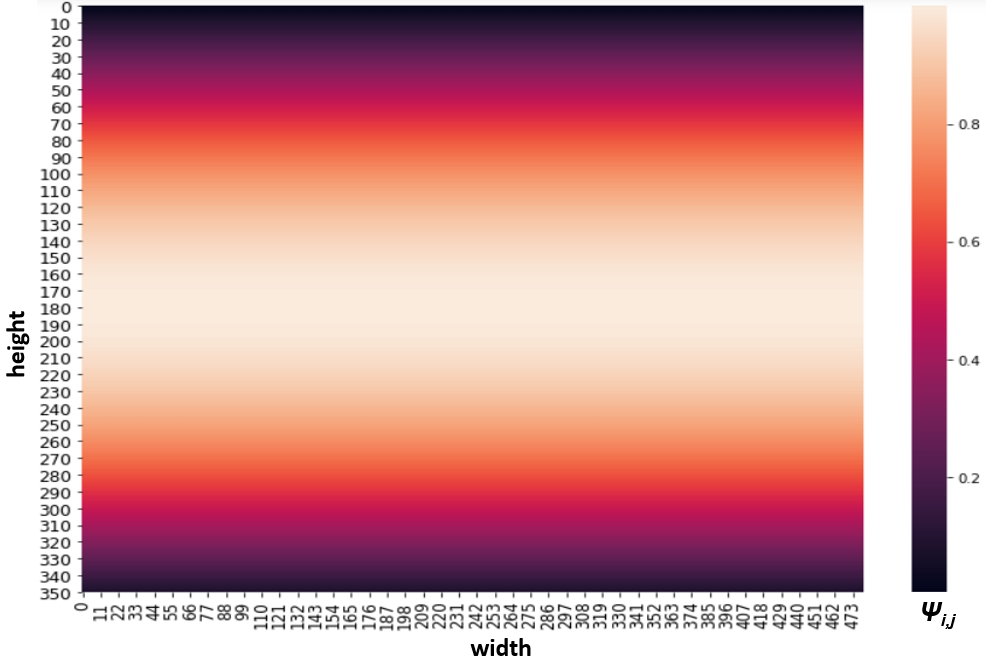}  
        \vspace{-1em}
         \caption{Heatmap to visualise the weight map ($\psi$) of size $480\times360$ pixels (same dimension as frames in 360VDS) based on $\psi_{i,j}$ definition from Eqn.(~\ref{eqn:loss}).}
         \label{fig:dist_map}
    \end{figure}

    Considering  distortion, the International Telecommunication Union (ITU)
    pioneered the use of distortion maps to be applied on conventional Peak Signal-to-Noise Ratio (PSNR) for more accurate quality evaluations~\cite{sun2016ahg8}. Following in these footsteps, we propose to use a weight map with a Smooth L1 loss to account for the distortion in {360\textdegree} videos during training to generate a super-resolver. To our knowledge, this is the first time a Weighted Spherically Smooth-L1 (WSS-L1) Loss function, shown in Eqn.~(\ref{eqn:loss}), has been used for this purpose. In Eqn.~(\ref{eqn:loss}), \textit{width} and \textit{height} represent the horizontal and vertical resolution of the generated High-Resolution (HR) output, which is the same as that of the Ground Truth (GT); and, the weights for a given row ($j$) in the distortion map $\psi$ remains same, and it only varies along the latitude. An example distortion map of size $480\times360$ pixels is visualised with the help of a heatmap in Fig.~\ref{fig:dist_map}. As seen in this figure, the use of a weight map allows the deep learning model to pay more importance to equatorial regions with higher weight values, the most focused area within the wide FoV of {360\textdegree} videos~\cite{lin2017tell,deng2021lau}. This can be used as a standard loss function for future research and related products specifically for {360\textdegree} signals in EquiRectangular Projection (ERP)  form.

     \begin{figure*}[!t]
         \centering
         \includegraphics[width = \textwidth]{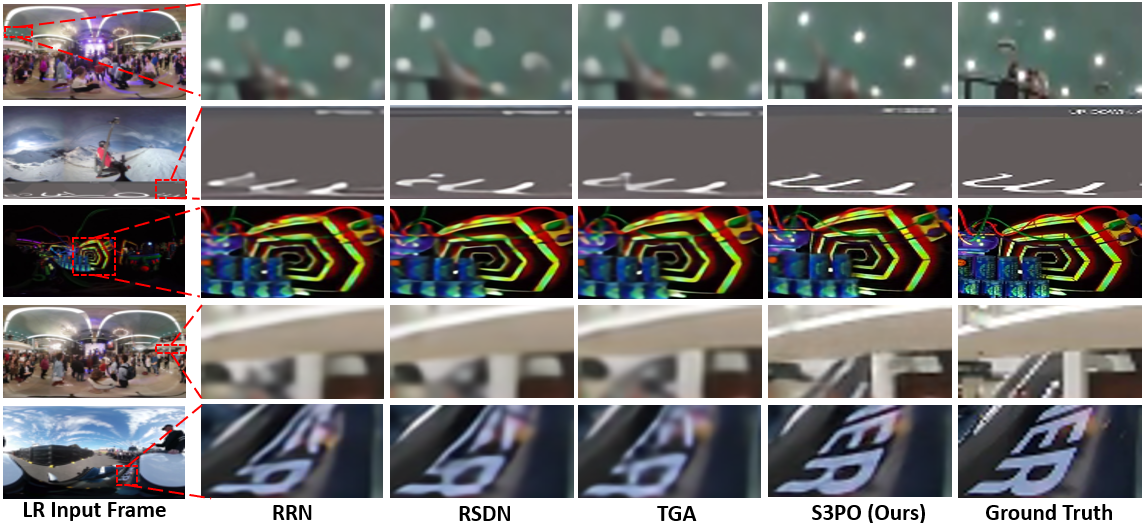}
         \vspace{-2em}
         \caption{Visually inspected qualitative performance comparison for super-resolution models using Blur Downsampling (BD). Super-resolved output for segments profiled from five different Low Resolution (LR) frames belonging to five different clips confirms S3PO's outstanding visual restoration ability in reference to the Ground Truth (GT). }
         \label{Fig:comparison}
         \vspace{-1em}
    \end{figure*}
         \begin{table*}[!t]
    
    \caption{{\bf Quantitative comparison using PSNR in dB, SSIM, WS-SSIM and Weighted Spherically-PSNR (WS-PSNR) in dB~\cite{sun2016ahg8} on the 360VDS Testset.} All results are computed in the luma (Y) channel. The {\color{red}red} and {\color{blue}blue} highlight the highest and second-highest values, respectively. \rev{Applicability indicates whether the model is applicable in offline or online settings. The \# Input Frames refers to the number of frames used as input to super-resolve a given target frame.} The column of Params(M) shows each model's complexity in millions of parameters. Degradation indicates whether the LR input used is generated using BI or BD degradation.}
    \centering
    \begin{tabular}{@{\hspace{0pt}} l @{\hspace{9pt}} c @{\hspace{9pt}} c @{\hspace{9pt}} c @{\hspace{9pt}} c @{\hspace{9pt}} c @{\hspace{9pt}} c @{\hspace{9pt}} c @{\hspace{9pt}} c @{\hspace{0pt}}}
    \hline
     Model Name & \rev{Applicability} & \rev{\# Input Frames} & Params(M) & Degradation &PSNR(Y) & SSIM(Y) & WS-PSNR(Y) & WS-SSIM(Y)\\
    
      \hline
       Bicubic & \rev{-}& \rev{-}& -& BI& 25.22 & 0.7436 & 24.46 & 0.7177\\
     DUF~\cite{DUF} & \rev{online, offline} & \rev{7} & 5.8&BI &26.46 &0.817 & 25.73& 0.7909\\
     RBPN~\cite{RBPN} & \rev{online, offline} & \rev{7} &12.2 &BI &{\color{blue}{27.14}} & \color{blue}{0.8214}& {\color{blue}{26.27}} &0.8004\\
     EDVR~\cite{EDVR} & \rev{online, offline} & \rev{7} & 3.3&BI & 26.85&0.8147 & 25.99 & 0.7946\\
    \rev{BasicVSR~\cite{BasicVSR}} & \rev{offline only} & \rev{all} & 6.3&BI & 27.05&{\color{red}{0.8227}} & 26.19 & {\color{blue}{0.8027}}\\
     \hline
     S3PO (Ours) &\rev{online, offline} & \rev{3 }& 8.89 & BI &{\color{red}{27.26}} & {\color{red}{0.8227}}& {\color{red}{26.32}} & {\color{red}{0.8030}} \\
     \hline\hline
     
     TGA~\cite{TGA} & \rev{online, offline} & \rev{7} & 5.8& BD &27.31 &0.8298 & {\color{blue}{26.39}} & 0.8095\\
     RSDN~\cite{RSDN} & \rev{online, offline} & \rev{2}& 6.2&BD &{\color{blue}{27.32}} &{\color{blue}{0.8313}} & 26.37 &{\color{blue}{0.8108}} \\
      RRN~\cite{RRN} & \rev{online, offline} & \rev{2} & 3.4& BD &26.96 &0.8214 & 26.07 & 0.8003\\
     \hline
      S3PO (Ours) &\rev{online, offline} & \rev{3} &  8.89& BD &{\color{red}{27.51}} & {\color{red}{0.8326}}& {\color{red}{26.52}} & {\color{red}{0.8120}} \\
       \hline
     \end{tabular}
      \label{tab:360VDS_result}
      \vspace{-1em}
     \end{table*}

    \subsection{Quantitative Evaluation}
    \label{sec:quant_eval}
    The performance of the proposed S3PO model, against other state-of-the-art conventional and {360\textdegree} specific VSR models, is evaluated on two test sets, namely the 360VDS -- test set with 45 clips and the MiG Panorama test set with four clips. For a fair comparison, and assessment with more recent conventional VSR models, we train and evaluate the S3PO model with two types of input degradation, namely Bicubic Interpolation (BI) and Blur Degradation (BD). To measure the quality of the generated super-resolved output, both conventional quality evaluation metrics (PSNR and SSIM), and {360\textdegree} specific quality evaluation metrics (Weighted Spherically -- PSNR (WS-PSNR)~\cite{sun2016ahg8} and WS-SSIM~\cite{zhou2018weighted}) are used.
    
    On average, across the 45 clips in the test set of 360VDS, S3PO performs best across all evaluation metrics and types of degradation as shown in Table~\ref{tab:360VDS_result}. The conventional VSR model used in this table uses the original degradation, as presented by the corresponding authors. RBPN~\cite{RBPN} and RSDN~\cite{RSDN} mostly perform second best to S3PO when degradation is BI and BD, respectively. \rev{For BD degradation, TGA~\cite{TGA} outperforms RSDN only on the WS-PSNR metric, and S3PO outperforms all the models on all metrics. Notably, on the BI degradation, BasicVSR~\cite{BasicVSR} trained on REDS~\cite{REDS} achieves higher SSIM and WS-SSIM scores than RBPN, resulting in the SSIM performance same as that of S3PO and second to the best performance on WS-SSIM. Nevertheless, S3PO's superior performance is observed over all other models for BI degradation as well.}
    
    Furthermore, the S3PO model outperforms all super-resolution models across all evaluation metrics in the MiG Panorama test set, as presented in Table~\ref{tab:mig_results}, while ~\rev{BasicVSR~\cite{BasicVSR}} performs second best to S3PO. In this case, all the models presented make use of BI degradation. This demonstrates the effective performance of the S3PO model in the quality enhancement of {360\textdegree} videos in different datasets representing varying input conditions. 

   \rev{ A further evaluation and comparison of S3PO and the best performing conventional VSR models -- BasicVSR~\cite{BasicVSR} -- is presented in Table~\ref{tab:hq_result}. The BasicVSR model trained on conventional high-resolution dataset REDS~\cite{REDS} outperforms most of the existing conventional VSR models in super-resolving high-resolution conventional videos such as the REDS test set and UDM10~\cite{UDM10} dataset. We evaluate the S3PO model and BasicVSR model on the newly created 360 Ultra HD dataset, which poses a high-resolution benchmarking challenge in {360\textdegree} video super-resolution. As shown in Table~\ref{tab:hq_result}, the  S3PO model outperforms  BasicVSR  across all evaluation measures for all clips. Despite BasicVSR being trained with an objective to super-resolve high-resolution videos, conventional VSR models such as BasicVSR are not suited well for the unique nature of {360\textdegree} videos. Thus, our omnidirectional VSR model -- S3PO -- outperforms its conventional counterparts despite being only trained to super-resolve LR of resolution $120\times90$ pixels to HR of resolution $480\times360$ pixels. This provides further evidence of S3PO's robustness and superior ability to super-resolve diverse ranges of {360\textdegree} videos in EquiRectangular Projection (ERP)  form.   
   }
    \subsection{Visual Comparison}
    \label{sec:qual_eval}
    
    The qualitative performance of the proposed S3PO model, when compared to that of other models, can be inspected visually as in Fig.~\ref{Fig:comparison}. For this analysis, models using BD downsampling to generate LR inputs are considered. The enhanced spatial profile for a given LR input is compared with the corresponding spatial profile's visual quality in the Ground Truth (GT). For each of the five sampled frames, representing five different clips in Fig.~\ref{Fig:comparison}, it is evident that S3PO restores finer details across segments from various sections within the input frame -- ranging from inner contents to edges. The visual details restored are improved significantly in reference to the GT when compared to other state-of-the-art super-resolution models. This further confirms the exceptional enhancement ability of S3PO resulting from the multi-fold benefits of targeted modelling and training.
      \begin{table*}[ht]
    
      \caption{{\bf Quantitative comparison using PSNR, SSIM, WS-PSNR and WS-SSIM on MiG Panorama Testset.} All results are computed in the luma (Y) channel. The {\color{red}red} and {\color{blue}blue} highlight the highest and second-highest values respectively.}
    \centering
    \begin{tabular}{@{\hspace{0pt}} l @{\hspace{9pt}} c @{\hspace{9pt}} c @{\hspace{9pt}} c @{\hspace{9pt}} c @{\hspace{9pt}} c @{\hspace{0pt}}}
    \hline
     Model Name & WS-PSNR(Y) & WS-SSIM(Y) & PSNR(Y) & SSIM(Y)\\
     \hline
     Bicubic & 28.81 & 0.7964 & 29.00 & 0.8121\\
     SR360~\cite{ozcinar2019super} & 29.40 &0.8123 & 29.69 & 0.8272 \\
     VSRnet~\cite{kappeler2016video} & 29.10 &0.8112 & 29.30 & 0.8241\\
     VESPCN~\cite{caballero2017real} &  29.70&0.8268 &29.97 & 0.8388\\
    FRVSR~\cite{FRVSR} & 29.63 &0.8266 & 29.88 & 0.8389 \\
    RBPN~\cite{RBPN} &  29.76& 0.8319&30.16 &0.8446 \\
     EDVR~\cite{EDVR} &  29.9&0.8358 &30.32 &0.8479 \\
     SMFN~\cite{liu2020single} &  {30.13}& {0.8381} & {30.56} & {0.8505}\\
      \rev{BasicVSR~\cite{BasicVSR}} &  {\color{blue}{30.22}} &{\color{blue}{0.8452}} & {\color{blue}{30.89}} & {\color{blue}{0.8559}}\\
     \hline
     S3PO (Ours) & {\color{red}{30.42}} & {\color{red}{0.8453}} & {\color{red}{31.16}} & {\color{red}{0.8565 }} \\
     \hline
     \end{tabular}
      \label{tab:mig_results}
     \end{table*}

\begin{table*}[ht]
    \caption{\textbf{Quantitative comparison using PSNR in dB, SSIM, WS-SSIM and WS-PSNR in dB on the high-resolution - 360 Ultra High-Definition (360UHD) test set.} All results are computed in the luma (Y) channel. The resolution is for ground truth and the predicted HR frames. The \textbf {blod} highlights the highest performance. }
    \centering
    \begin{tabular}{@{\hspace{0pt}} l @{\hspace{9pt}} l |@{\hspace{3pt}} c @{\hspace{3pt}} c | @{\hspace{3pt}} c  @{\hspace{3pt}} c | @{\hspace{3pt}} c @{\hspace{3pt}} c |
    @{\hspace{3pt}} c @{\hspace{3pt}} c  @{\hspace{1pt}}}
    \hline
     Clip Name & Resolution & \multicolumn{2}{c}{PSNR(Y)} & \multicolumn{2}{c}{SSIM(Y)} & \multicolumn{2}{c}{WS-PSNR(Y)} & \multicolumn{2}{c}{WS-SSIM(Y)} \\
    \hline
     & & S3PO & BasicVSR & S3PO & BasicVSR & S3PO & BasicVSR& S3PO & BasicVSR \\
     \cline{3-10}
     Skiing & $1280\times720$& {\bf{32.86}} &32.80 & {\bf{0.9193}}&0.9161& {\bf{31.68}}&31.66& {\bf{0.9151}}&0.9126 \\
     Sunset & $1280\times720$ &\textbf{39.15} &38.82& \textbf{0.9264}&0.9224&\textbf{39.29}&38.94&\textbf{0.9310}&0.9269 \\
     Parade & $1280\times720$ & \textbf{24.62} &24.24& \textbf{0.8541}&0.8454&\textbf{24.10}&23.72&\textbf{0.8399}&0.8304 \\
     Cafeteria & $2560\times1280$& \textbf{37.13}&37.03&\textbf{0.9776}&0.9770&\textbf{35.41}&35.34&\textbf{0.9715}&0.9710\\
    Street & $3840\times1920$&\textbf{38.17} &37.65&\textbf{0.9771}&0.9741&\textbf{36.81}&36.24&\textbf{0.9720}&0.9680\\
    Welder & $3840\times1920$&\textbf{34.00} &33.52&\textbf{0.9432}&0.9380&\textbf{33.41}&32.98&\textbf{0.9425}&0.9373\\
    Baboon & $4096\times2048$& \textbf{30.00}&29.71&\textbf{0.8550}&0.8471&\textbf{30.32}&30.04&\textbf{0.8744}&0.8668\\
    Birds & $4096\times2048$&\textbf{26.17} &25.69&\textbf{0.7606}&0.7436&\textbf{25.79}&25.28&\textbf{0.7381}& 0.7189\\
       \hline
\end{tabular}

\label{tab:hq_result}
\vspace{-1em}
\end{table*}
       
 \begin{figure}[ht]
 \vspace{-1em}
         \centering
         \includegraphics[width = 0.2\textwidth]{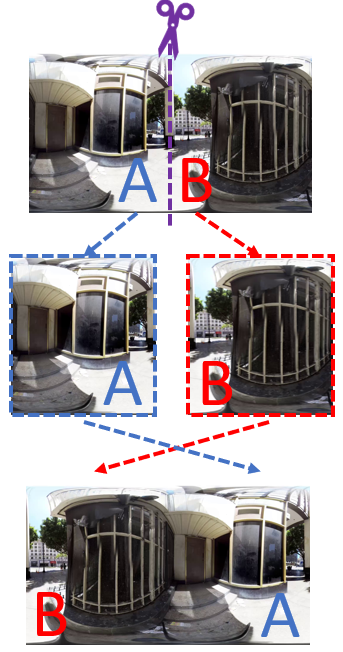} 
         \vspace{-1em}
         \caption{Illustration of steps in cyclic treatment applied to an ERP frame considering horizontal panoramic continuity of the scenes in {360\textdegree} videos.}
         \label{cyclic_treat}
         \vspace{-1em}
    \end{figure}

     \begin{table}[ht]
     \caption{{\bf Impact of cyclic treatment as discussed in Sec.~\ref{sec:horizontal_cyclicity}. }}
    \centering
    \begin{tabular}{@{\hspace{0pt}} c @{\hspace{4pt}} c @{\hspace{9pt}} c @{\hspace{9pt}} c @{\hspace{9pt}} c @{\hspace{9pt}} c  @{\hspace{0pt}}}
    \hline
     Model & Dataset & PSNR(Y) & SSIM(Y) & WS-PSNR(Y) & WS-SSIM(Y)\\
     \hline
    S3PO  & 360VDS &27.2613 & 0.8227& 26.3195 & 0.8030\\
     S3PO-cyc & 360VDS & {\bf{27.2618}} & {\bf{0.8227}} & {\bf{26.3236}} & {\bf{0.8030}} \\
     \hline
     S3PO  & MiG & 30.4203 & {\bf{0.8453}} & 31.1612& 0.8565\\
     S3PO-cyc & MiG & {\bf{30.4330}} & 0.8452 & {\bf{31.1925}} & {\bf{0.8566 }} \\
     \hline
     \end{tabular}
 \label{tab:cyclic}
 \vspace{-2em}
 \end{table}

\subsection{Horizontal Cyclicity in ERP Frames}
\label{sec:horizontal_cyclicity}
The horizontal cyclicity in EquiRectangular Projection (ERP)  frames of {360\textdegree} videos refers to the continuous scene that emerges when the left and right edges of the frame are stitched together, creating a seamless panoramic view~\cite{horizontal_cyc,ai2022deep}. This property can lead to the loss of finer details and degradation of ERP frames when super-resolution techniques are applied, as the model may not be able to effectively capture the complete content of the scene across both edges.

To address this issue, a cyclic treatment methodology is considered, as depicted in Fig.~\ref{cyclic_treat}. This method takes advantage of the geometric properties of the equirectangular projection by dividing a frame into two equal parts separated along the vertical axis, followed by stitching of the swapped left and right parts. This places the edge contents in the middle of the frame as part of a continuous scene, giving the model the opportunity to extract meaningful features for all contents across the frame, regardless of their location. The model is trained to extract features from both the original equirectangular frame and the frame subjected to the cyclic treatment, ensuring that each object or content is fed to the model as part of a continuous scene at least once.

The S3PO model is fine-tuned to extract features from the two variations of given input frames. The two sets of extracted features from the {360\textdegree} feature extractor are added element-wise at the co-located positions across the two feature sets. The feature obtained in this way is then propagated for refinement without further methodological changes. With this approach, no architectural complexity is introduced, keeping the structure and size of the model constant.

As shown in Table~\ref{tab:cyclic}, the cyclic treatment discussed above results in an improvement of PSNR and WS-PSNR results. S3PO-cyclic can be a useful variant for applications where higher pixel accuracy is desirable. At the same time, this also confirms that the proposed S3PO model with no cyclic overhead is adequately robust to the uniqueness of {360\textdegree} videos.

\section{Ablation Study}
\label{sec:ab_study}
The multi-fold strategic design choices that jointly lead to a superior super-resolution performance from the proposed S3PO model while also making it robust to the challenges posed by distortions in EquiRectangular Projections (ERPs) frames are studied in a two-staged ablation study presented in this section. For each ablation study, only one factor is investigated at a time while keeping the network size mostly similar in order to make a fair and factual comparison. All the models considered in this study are trained to super-resolve LR inputs generated from blur degradation.
\begin{figure*}[hb]
 \centering
 \includegraphics[width = 0.85\textwidth]{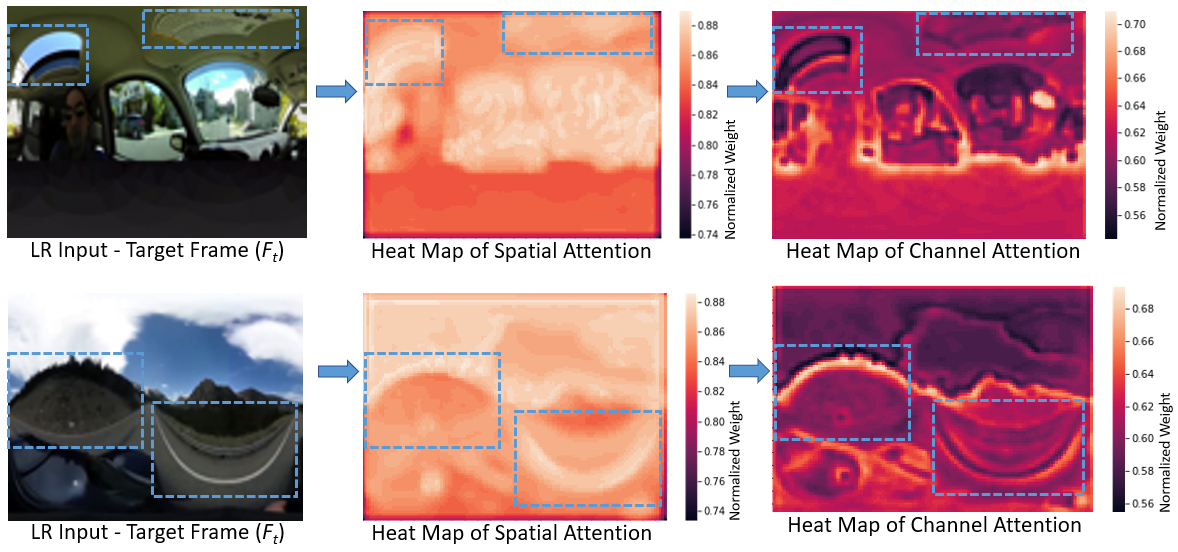}   
  \vspace{-0.8em}
 \caption{Heatmaps of attention weights obtained from the attention mechanism within {360\textdegree} Feature Extractor illustrating a strong correlation between learnt attention weights and the distorted regions of ERP frames (highlighted by bounding boxes). The heat maps of spatial attention weights help distinguish the varied degree of attention paid to the spatially distorted objects, while both spatial and channel attention weights ensure that the distorted nature is preserved in the learnt weights.}
 \label{Fig:attention_map}
\end{figure*}

\subsection{{360\textdegree}-specific Characteristics}

To understand the effectiveness of the proposed {360\textdegree} feature extractor, we replace it with a conventional MEMC-based alignment step by making use of the SpyNet~\cite{spynet} flow estimation model. Removing the {360\textdegree} feature extractor and replacing it with the pre-trained SpyNet model leads to a decrease in super-resolution performance across all evaluation metrics denoted by {\it'w/o {360\textdegree} Feature Extractor'} in Table~\ref{tab:attention}. This signifies the superiority of the proposed feature extraction mechanism over the conventional alignment step in capturing coherent spatiotemporal details for {360\textdegree} video super-resolution. 

Furthermore, the attention mechanism within the {360\textdegree} feature extractor is expected to allow the S3PO model to learn to pay varied attention to distorted spatial regions and diverse temporal contexts present in the given ERP frames. The results, as shown in Table~\ref{tab:attention}, reveal that the removal of the attention mechanism leads to a decrement in all the performance measures denoted with {\it'w/o Attention'} in Table~\ref{tab:attention}. This confirms that 
the attention mechanism helps the network to better capture and weigh the contribution of different features to the final representation. As shown in Fig.~\ref{Fig:attention_map}, the visualization of the attention map illustrates a strong relationship between the distorted nature of the objects/contents and corresponding attention maps. Evidently, the distortions caused to objects and contents in polar regions of LR inputs due to the equirectangular projections in ERP frames are represented well by the learnt attention weights. The heat maps of spatial attention weights help distinguish the varied degree of attention paid to the spatially distorted objects, while both spatial and channel attention weights ensure that the distorted nature is preserved in the learnt weights. These illustrations further showcase the learnt ability of the attention mechanism to preserve and account for the distortion present in ERP frames. 
\begin{figure*}[hb]
 \centering
 \includegraphics[width = 0.85\textwidth]{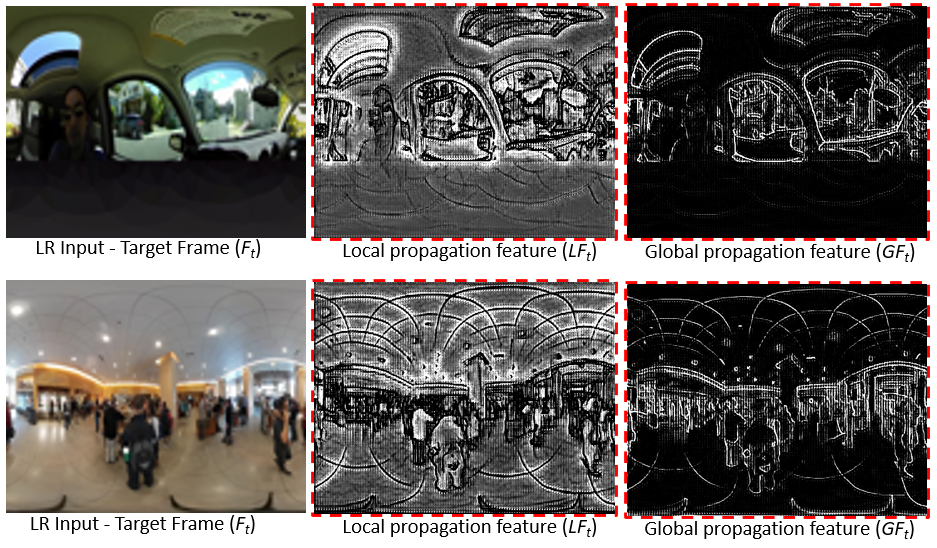}
 \vspace{-0.8em}
 \caption{Final features obtained from two ducts of S3PO model illustrating different frequency details represented mutually by the ducts for a given LR input.}
 \label{Fig:dual_feats}
\end{figure*}
 \subsection{Information Propagation and Domain Adaptation}
\begin{table}[ht]
      \caption{{\bf S3PO with and without {360\textdegree}-specific considerations.}The Params represents parameters in millions. All the results are computed in the Y channel and for BD degradation.}
    \centering
    \begin{tabular}{@{\hspace{0pt}} c @{\hspace{3pt}} c @{\hspace{3pt}} c @{\hspace{3pt}} c @{\hspace{3pt}} c @{\hspace{3pt}} c  @{\hspace{0pt}}}
    \hline
     S3PO & Params  & PSNR & SSIM & WS-PSNR & WS-SSIM\\
     \hline
      w/o {360\textdegree} Feat. Extractor & 8.07 &27.16&0.8225&26.22&0.8014\\
     w/o Weighted Loss & 8.89 & 27.46 &0.8317 & 26.46& 0.8105\\
      w/o Attention  & 7.22 & 27.49 &0.8322 & 26.48& 0.8111\\
    Full-model & 8.89 & {\bf 27.51} & {\bf 0.8326}& {\bf 26.52} & {\bf 0.8120} \\
     \hline
     \end{tabular}
 \label{tab:attention}
 \end{table}
\begin{table}[ht!]
      \caption{{\bf Impact of WSS-L1 Loss $\beta$ values(see eqn.~\ref{eqn:loss}) on S3PO results.}}
    \centering
    \begin{tabular}{@{\hspace{0pt}} c @{\hspace{9pt}} c @{\hspace{9pt}} c @{\hspace{9pt}} c @{\hspace{9pt}} c  @{\hspace{0pt}}}
    \hline
     $\beta$ & PSNR(Y) & SSIM(Y) & WS-PSNR(Y) & WS-SSIM(Y)\\
     \hline
     0.5 & 27.50&0.8324 & 26.51 & 0.8118\\
     1 (Default Setting) & {\bf 27.51} & {\bf 0.8326}& {\bf 26.52} & {\bf 0.8120}\\
     2 & 27.49 & 0.8322& 26.50 & 0.8116\\
     \hline
     \end{tabular}
    \label{tab:beta_values}
 \end{table}

The S3PO model is trained to optimize the proposed weighted spherically smooth-l1 (WSS-L1) loss as seen in eqn.~\ref{eqn:loss}. To understand the effectiveness of the WSS-L1 loss for {360\textdegree} video super-resolution task, we replace WSS-L1 loss with conventional Smooth-L1 loss. This resulted in a deterioration in performance denoted with {\it'w/o Weighted Loss'} in Table~\ref{tab:attention}; thus affirming the improved ability of the S3PO model in dealing with the uniqueness of ERP frames and preserving those characteristics during  super-resolution.

The balance between the L1 loss and L2 loss components in WSS-L1 loss is controlled by the $\beta$ parameter as seen in Eqn.~\ref{eqn:loss}.  L1 loss measures the absolute difference between the predicted and ground truth values, while  L2 loss measures the squared difference. A larger $\beta$ value places more weight on the L2 loss component and is more sensitive to larger differences and outliers, while a smaller $\beta$ value places more weight on the L1 loss component and is more sensitive to smaller differences and is likely to produce erratic gradients and an unstable convergence process. 

To study the impact of $\beta$ in balancing the contributions of L1 and L2 norms in WSS-L1 loss, the default $\beta=1$ value was doubled and halved. Reducing the value of the $\beta$ parameter by half places more weight on the L1 loss component and less weight on the L2 loss component, while doubling the value does the exact opposite. As shown in Table~\ref{tab:beta_values}, changes up and down to $\beta$ lead to inferior performance compared to $\beta = 1$. The superior results across all evaluation metrics indicate that $\beta=1$ is the optimal choice that maintains the desired balance between L1 and L2 norms in the proposed WSS-L1 loss. Thus $\beta=1$ is used as the default parameter setting for all S3PO training presented in this paper.

 \begin{table}[ht] 
      \caption{{\bf Impact of recurrent residue, mutual information exchange and domain adaptation on S3PO results.}The Params represent parameters in millions. All the results are computed in the Y channel and for BD degradation.}
    \centering
    \begin{tabular}{@{\hspace{4pt}} c @{\hspace{4pt}} c @{\hspace{4pt}} c @{\hspace{4pt}} c @{\hspace{4pt}} c @{\hspace{4pt}} c  @{\hspace{4pt}}}
    \hline
     S3PO & Params & PSNR & SSIM & WS-PSNR & WS-SSIM\\
     \hline
     w/o Hidden-State  & 8.47&27.30&0.8266 & 26.32& 0.8054\\
     w/o Mutual Exchange & 8.89&27.37&0.8287 &26.40 &0.8080 \\
     w/o Domain Adaptation  & 8.89&27.46 &0.8317 & 26.46& 0.8105\\
     Full-model & 8.89 &{\bf 27.51} & {\bf 0.8326}& {\bf 26.52} & {\bf 0.8120} \\
     \hline
     \end{tabular}
 \label{tab:domain_adapt}
 \end{table}

The use of recurrent residues is aimed to enable the S3PO model to capture temporal dependencies between adjacent frames in a video, thereby improving the quality of the super-resolved output. At the same time, the use of a dual duct architecture, as seen in Fig~\ref{Fig:res_block}, allows for the mutual exchange between local information (induced by {360\textdegree} feature extractor) and global information (generated by recurrent residues) propagation. This means that information from both fine and coarse levels, as shown in Fig.~\ref{Fig:dual_feats}, can be passed between different processing stages. This allows for the effective use of both local and global information, leading to a more accurate representation of the high-resolution frames.

The empirical results presented in the Table~\ref{tab:domain_adapt} show that the removal of the recurrent residues (hidden state $h_{t-1}$ as seen in Eqn.(~\ref{eq:1})) and mutual exchange (as seen in Eqns.~(\ref{eq:6}-\ref{eq:5}) between local and global information propagation leads to a significant reduction for all quality measures. This provides evidence of the practical advantages of the use of these techniques in improving the accuracy and quality of the super-resolved output in the context of {360\textdegree} videos.

Similarly, the S3PO model without domain adaptation underperforms compared to the version with adaptation in all four evaluation metrics, as shown in Table~\ref{tab:domain_adapt}. The benefits of domain adaptation in this context of video super-resolution can be attributed to the improved ability of the S3PO model to generalize to new, unseen data. By fine-tuning the S3PO model on the {360\textdegree} video data set, the model is better able to handle the unique challenges posed by this type of data while leveraging from the base initialization of super-resolution task from the conventional domain.

\section{Conclusion and Future Work}

The applicability of conventional Video Super-Resolution (VSR) models, with satisfactory outcomes on {360\textdegree} videos, is demonstrated in this research. To ensure diverse training and test conditions, desirable for low-level computer vision systems specific to {360\textdegree} multimedia, a novel dataset is assembled and described. Conventional VSR models are applicable to omnidirectional videos because the EquiRectangular Projection (ERP) frames are similar in format to conventional video frames. Nevertheless, the data within an ERP frame is unique and different from the conventional video frame because of the distortion present along the vertical axis \rev{and cyclic continuity along the horizontal aixs} in {360\textdegree} videos. 

Accounting for the ERP-specific properties, a novel {360\textdegree} VSR model is proposed (the S3PO model) with an ERP-specific architecture, feature extractor and optimiser. The empirical evaluation and ablation study confirm the superiority of the super-resolution performance of the S3PO model provisioned by the combined benefit of 360\textdegree-content specific architectural sub-components, training with domain adaptation and optimisation with distortion-aware loss. 
The  S3PO model does not incorporate conventional VSR steps, such as alignment; nonetheless, it outperforms the state-of-the-art super-resolution models, including those that use alignment. 
\balance

The S3PO model and the 360VDS dataset help define new opportunities for future {360\textdegree} multimedia research. The application of implicit and explicit alignment techniques as extension work can be further studied, with appropriately attenuated alignments to account for the distortion \rev{and cyclicity} in EquiRectangular Projections (ERPs) frames. Additionally,  the impact on Quality-of-Experience (QoE) resulting due to quality enhancement using S3PO can be conducted to better understand how the model \rev{impacts  the user perception and consumption} of {360\textdegree} multimedia.

\bibliographystyle{IEEEtran}
\bibliography{reference}

\begin{thebibliography}{10}
\providecommand{\url}[1]{#1}
\csname url@samestyle\endcsname
\providecommand{\newblock}{\relax}
\providecommand{\bibinfo}[2]{#2}
\providecommand{\BIBentrySTDinterwordspacing}{\spaceskip=0pt\relax}
\providecommand{\BIBentryALTinterwordstretchfactor}{4}
\providecommand{\BIBentryALTinterwordspacing}{\spaceskip=\fontdimen2\font plus
\BIBentryALTinterwordstretchfactor\fontdimen3\font minus
  \fontdimen4\font\relax}
\providecommand{\BIBforeignlanguage}[2]{{%
\expandafter\ifx\csname l@#1\endcsname\relax
\typeout{** WARNING: IEEEtran.bst: No hyphenation pattern has been}%
\typeout{** loaded for the language `#1'. Using the pattern for}%
\typeout{** the default language instead.}%
\else
\language=\csname l@#1\endcsname
\fi
#2}}
\providecommand{\BIBdecl}{\relax}
\BIBdecl

\bibitem{maniotis2019tile}
P.~Maniotis, E.~Bourtsoulatze, and N.~Thomos, ``Tile-based joint caching and
  delivery of 360 videos in heterogeneous networks,'' \emph{IEEE Transactions
  on Multimedia}, vol.~22, no.~9, pp. 2382--2395, 2019.

\bibitem{yaqoob2020survey}
A.~Yaqoob, T.~Bi, and G.-M. Muntean, ``A survey on adaptive 360 video
  streaming: solutions, challenges and opportunities,'' \emph{IEEE
  Communications Surveys \& Tutorials}, vol.~22, no.~4, pp. 2801--2838, 2020.

\bibitem{zhou2019video}
Y.~Zhou, L.~Tian, C.~Zhu, X.~Jin, and Y.~Sun, ``Video coding optimization for
  virtual reality 360-degree source,'' \emph{IEEE Journal of Selected Topics in
  Signal Processing}, vol.~14, no.~1, pp. 118--129, 2019.

\bibitem{10.1145/3097895.3097896}
\BIBentryALTinterwordspacing
S.~Afzal, J.~Chen, and K.~K. Ramakrishnan, ``Characterization of 360-degree
  videos,'' in \emph{Proceedings of the Workshop on Virtual Reality and
  Augmented Reality Network}, ser. VR/AR Network '17.\hskip 1em plus 0.5em
  minus 0.4em\relax New York, NY, USA: Association for Computing Machinery,
  2017, p. 1–6. [Online]. Available:
  \url{https://doi-org.ezproxy-f.deakin.edu.au/10.1145/3097895.3097896}
\BIBentrySTDinterwordspacing

\bibitem{9155477}
M.~Dasari, A.~Bhattacharya, S.~Vargas, P.~Sahu, A.~Balasubramanian, and S.~R.
  Das, ``Streaming 360-degree videos using super-resolution,'' in \emph{IEEE
  INFOCOM 2020 - IEEE Conference on Computer Communications}, 2020, pp.
  1977--1986.

\bibitem{8329628}
M.~S. Elbamby, C.~Perfecto, M.~Bennis, and K.~Doppler, ``Toward low-latency and
  ultra-reliable virtual reality,'' \emph{IEEE Network}, vol.~32, no.~2, pp.
  78--84, 2018.

\bibitem{RBPN}
M.~Haris, G.~Shakhnarovich, and N.~Ukita, ``Recurrent back-projection network
  for video super-resolution,'' in \emph{Proceedings of the IEEE/CVF Conference
  on Computer Vision and Pattern Recognition}, 2019, pp. 3897--3906.

\bibitem{EDVR}
X.~Wang, K.~C. Chan, K.~Yu, C.~Dong, and C.~Change~Loy, ``Edvr: Video
  restoration with enhanced deformable convolutional networks,'' in
  \emph{Proceedings of the IEEE/CVF Conference on Computer Vision and Pattern
  Recognition Workshops}, 2019.

\bibitem{RSDN}
T.~Isobe, X.~Jia, S.~Gu, S.~Li, S.~Wang, and Q.~Tian, ``Video super-resolution
  with recurrent structure-detail network,'' in \emph{Computer Vision -- ECCV
  2020}, A.~Vedaldi, H.~Bischof, T.~Brox, and J.-M. Frahm, Eds.\hskip 1em plus
  0.5em minus 0.4em\relax Cham: Springer International Publishing, 2020, pp.
  645--660.

\bibitem{BasicVSR}
K.~C. Chan, X.~Wang, K.~Yu, C.~Dong, and C.~C. Loy, ``Basicvsr: The search for
  essential components in video super-resolution and beyond,'' in
  \emph{Proceedings of the IEEE/CVF Conference on Computer Vision and Pattern
  Recognition (CVPR)}, June 2021, pp. 4947--4956.

\bibitem{TGA}
T.~Isobe, S.~Li, X.~Jia, S.~Yuan, G.~Slabaugh, C.~Xu, Y.-L. Li, S.~Wang, and
  Q.~Tian, ``Video super-resolution with temporal group attention,'' in
  \emph{Proceedings of the IEEE/CVF Conference on Computer Vision and Pattern
  Recognition (CVPR)}, June 2020.

\bibitem{deformConv}
J.~Dai, H.~Qi, Y.~Xiong, Y.~Li, G.~Zhang, H.~Hu, and Y.~Wei, ``Deformable
  convolutional networks,'' in \emph{Proceedings of the IEEE international
  conference on computer vision}, 2017, pp. 764--773.

\bibitem{DUF}
Y.~Jo, S.~W. Oh, J.~Kang, and S.~J. Kim, ``Deep video super-resolution network
  using dynamic upsampling filters without explicit motion compensation,'' in
  \emph{Proceedings of the IEEE conference on computer vision and pattern
  recognition}, 2018, pp. 3224--3232.

\bibitem{FeatGAN}
A.~Lucas, S.~Lopez-Tapia, R.~Molina, and A.~K. Katsaggelos, ``Generative
  adversarial networks and perceptual losses for video super-resolution,''
  \emph{IEEE Trans Image Process}, vol.~28, no.~7, pp. 3312--3327, 2019.

\bibitem{liu2020single}
H.~Liu, Z.~Ruan, C.~Fang, P.~Zhao, F.~Shang, Y.~Liu, and L.~Wang, ``A single
  frame and multi-frame joint network for 360-degree panorama video
  super-resolution,'' \emph{arXiv preprint arXiv:2008.10320}, 2020.

\bibitem{RRN}
T.~Isobe, F.~Zhu, X.~Jia, and S.~Wang, ``Revisiting temporal modeling for video
  super-resolution,'' \emph{arXiv preprint arXiv:2008.05765}, 2020.

\bibitem{R2D2}
\BIBentryALTinterwordspacing
A.~{Agrahari Baniya}., T.~Lee., P.~Eklund, S.~Aryal, and A.~Robles-Kelly,
  ``{Online Video Super-Resolution using Unidirectional Recurrent Model},'' 11
  2022. [Online]. Available:
  \url{https://www.techrxiv.org/articles/preprint/Online_Video_Super-Resolution_using_Unidirectional_Recurrent_Model/21500235}
\BIBentrySTDinterwordspacing

\bibitem{basic++}
K.~C. Chan, S.~Zhou, X.~Xu, and C.~C. Loy, ``Basicvsr++: Improving video
  super-resolution with enhanced propagation and alignment,'' in
  \emph{Proceedings of the IEEE/CVF conference on computer vision and pattern
  recognition}, 2022, pp. 5972--5981.

\bibitem{yi2021omniscient}
P.~Yi, Z.~Wang, K.~Jiang, J.~Jiang, T.~Lu, X.~Tian, and J.~Ma, ``Omniscient
  video super-resolution,'' in \emph{Proceedings of the IEEE/CVF International
  Conference on Computer Vision}, 2021, pp. 4429--4438.

\bibitem{fakour2018360}
V.~Fakour-Sevom, E.~Guldogan, and J.-K. K{\"a}m{\"a}r{\"a}inen, ``360 panorama
  super-resolution using deep convolutional networks,'' in \emph{Int. Conf. on
  Computer Vision Theory and Applications (VISAPP)}, vol.~1, 2018.

\bibitem{8901764}
C.~Ozcinar, A.~Rana, and A.~Smolic, ``Super-resolution of omnidirectional
  images using adversarial learning,'' in \emph{2019 IEEE 21st International
  Workshop on Multimedia Signal Processing (MMSP)}, 2019, pp. 1--6.

\bibitem{9506233}
A.~Nishiyama, S.~Ikehata, and K.~Aizawa, ``360° single image super resolution
  via distortion-aware network and distorted perspective images,'' in
  \emph{2021 IEEE International Conference on Image Processing (ICIP)}, 2021,
  pp. 1829--1833.

\bibitem{dong2015image}
C.~Dong, C.~C. Loy, K.~He, and X.~Tang, ``Image super-resolution using deep
  convolutional networks,'' \emph{IEEE transactions on pattern analysis and
  machine intelligence}, vol.~38, no.~2, pp. 295--307, 2015.

\bibitem{xiao2012recognizing}
J.~Xiao, K.~A. Ehinger, A.~Oliva, and A.~Torralba, ``Recognizing scene
  viewpoint using panoramic place representation,'' in \emph{2012 IEEE
  Conference on Computer Vision and Pattern Recognition}.\hskip 1em plus 0.5em
  minus 0.4em\relax IEEE, 2012, pp. 2695--2702.

\bibitem{dasari2020streaming}
M.~Dasari, A.~Bhattacharya, S.~Vargas, P.~Sahu, A.~Balasubramanian, and S.~R.
  Das, ``Streaming 360-degree videos using super-resolution,'' in \emph{IEEE
  INFOCOM 2020-IEEE Conference on Computer Communications}.\hskip 1em plus
  0.5em minus 0.4em\relax IEEE, 2020, pp. 1977--1986.

\bibitem{hore2010image}
A.~Hore and D.~Ziou, ``Image quality metrics: Psnr vs. ssim,'' in \emph{2010
  20th international conference on pattern recognition}.\hskip 1em plus 0.5em
  minus 0.4em\relax IEEE, 2010, pp. 2366--2369.

\bibitem{zhou2018weighted}
Y.~Zhou, M.~Yu, H.~Ma, H.~Shao, and G.~Jiang, ``Weighted-to-spherically-uniform
  ssim objective quality evaluation for panoramic video,'' in \emph{2018 14th
  IEEE International Conference on Signal Processing (ICSP)}.\hskip 1em plus
  0.5em minus 0.4em\relax IEEE, 2018, pp. 54--57.

\bibitem{spynet}
A.~Ranjan and M.~J. Black, ``Optical flow estimation using a spatial pyramid
  network,'' in \emph{Proceedings of the IEEE conference on computer vision and
  pattern recognition}, 2017, pp. 4161--4170.

\bibitem{10.1145/3411764.3445499}
\BIBentryALTinterwordspacing
P.~Bala, I.~Oakley, V.~Nisi, and N.~J. Nunes, ``Dynamic field of view
  restriction in 360° video: Aligning optical flow and visual slam to mitigate
  vims,'' in \emph{Proceedings of the 2021 CHI Conference on Human Factors in
  Computing Systems}, ser. CHI '21.\hskip 1em plus 0.5em minus 0.4em\relax New
  York, NY, USA: Association for Computing Machinery, 2021. [Online].
  Available: \url{https://doi.org/10.1145/3411764.3445499}
\BIBentrySTDinterwordspacing

\bibitem{bhandari2021revisiting}
K.~Bhandari, Z.~Zong, and Y.~Yan, ``Revisiting optical flow estimation in 360
  videos,'' in \emph{2020 25th International Conference on Pattern Recognition
  (ICPR)}.\hskip 1em plus 0.5em minus 0.4em\relax IEEE, 2021, pp. 8196--8203.

\bibitem{shi2022panoflow}
H.~Shi, Y.~Zhou, K.~Yang, Y.~Ye, X.~Yin, Z.~Yin, S.~Meng, and K.~Wang,
  ``Panoflow: Learning optical flow for panoramic images,'' \emph{arXiv
  preprint arXiv:2202.13388}, 2022.

\bibitem{lucas2019generative}
A.~Lucas, S.~Lopez-Tapia, R.~Molina, and A.~K. Katsaggelos, ``Generative
  adversarial networks and perceptual losses for video super-resolution,''
  \emph{IEEE Transactions on Image Processing}, vol.~28, no.~7, pp. 3312--3327,
  2019.

\bibitem{woo2018cbam}
S.~Woo, J.~Park, J.-Y. Lee, and I.~S. Kweon, ``Cbam: Convolutional block
  attention module,'' in \emph{Proceedings of the European conference on
  computer vision (ECCV)}, 2018, pp. 3--19.

\bibitem{li2018bridge}
C.~Li, M.~Xu, X.~Du, and Z.~Wang, ``Bridge the gap between vqa and human
  behavior on omnidirectional video: A large-scale dataset and a deep learning
  model,'' in \emph{Proceedings of the 26th ACM international conference on
  Multimedia}, 2018, pp. 932--940.

\bibitem{su2018learning}
Y.-C. Su and K.~Grauman, ``Learning compressible 360° video isomers,'' in
  \emph{Proceedings of the IEEE Conference on Computer Vision and Pattern
  Recognition}, 2018, pp. 7824--7833.

\bibitem{10.1145/3304109.3325820}
\BIBentryALTinterwordspacing
A.~Nguyen and Z.~Yan, ``A saliency dataset for 360-degree videos,'' in
  \emph{Proceedings of the 10th ACM Multimedia Systems Conference}, ser. MMSys
  '19.\hskip 1em plus 0.5em minus 0.4em\relax New York, NY, USA: Association
  for Computing Machinery, 2019, p. 279–284. [Online]. Available:
  \url{https://doi-org.ezproxy-b.deakin.edu.au/10.1145/3304109.3325820}
\BIBentrySTDinterwordspacing

\bibitem{chiariotti2021survey}
F.~Chiariotti, ``A survey on 360-degree video: Coding, quality of experience
  and streaming,'' \emph{Computer Communications}, vol. 177, pp. 133--155,
  2021.

\bibitem{xu2020state}
M.~Xu, C.~Li, S.~Zhang, and P.~Le~Callet, ``State-of-the-art in 360 video/image
  processing: Perception, assessment and compression,'' \emph{IEEE Journal of
  Selected Topics in Signal Processing}, vol.~14, no.~1, pp. 5--26, 2020.

\bibitem{10.1145/3304109.3325812}
\BIBentryALTinterwordspacing
A.~T. Nasrabadi, A.~Samiei, A.~Mahzari, R.~P. McMahan, R.~Prakash, M.~C.~Q.
  Farias, and M.~M. Carvalho, ``A taxonomy and dataset for 360° videos,'' in
  \emph{Proceedings of the 10th ACM Multimedia Systems Conference}, ser. MMSys
  '19.\hskip 1em plus 0.5em minus 0.4em\relax New York, NY, USA: Association
  for Computing Machinery, 2019, p. 273–278. [Online]. Available:
  \url{https://doi.org/10.1145/3304109.3325812}
\BIBentrySTDinterwordspacing

\bibitem{miller2020personal}
M.~R. Miller, F.~Herrera, H.~Jun, J.~A. Landay, and J.~N. Bailenson, ``Personal
  identifiability of user tracking data during observation of 360-degree vr
  video,'' \emph{Scientific Reports}, vol.~10, no.~1, pp. 1--10, 2020.

\bibitem{pyscene}
\BIBentryALTinterwordspacing
B.~Castellano. (2022) Pyscenedetect. [Online]. Available:
  \url{http://scenedetect.com/en/latest/}
\BIBentrySTDinterwordspacing

\bibitem{itu1999subjective}
T.~Installations and L.~Line, ``Subjective video quality assessment methods for
  multimedia applications,'' \emph{Networks}, vol. 910, no.~37, p.~5, 1999.

\bibitem{vimeo}
T.~Xue, B.~Chen, J.~Wu, D.~Wei, and W.~T. Freeman, ``Video enhancement with
  task-oriented flow,'' \emph{International Journal of Computer Vision}, vol.
  127, no.~8, pp. 1106--1125, 2019.

\bibitem{REDS}
S.~Nah, S.~Baik, S.~Hong, G.~Moon, S.~Son, R.~Timofte, and K.~Mu~Lee, ``Ntire
  2019 challenge on video deblurring and super-resolution: Dataset and study,''
  in \emph{Proceedings of the IEEE/CVF Conference on Computer Vision and
  Pattern Recognition (CVPR) Workshops}, June 2019.

\bibitem{UDM10}
P.~Yi, Z.~Wang, K.~Jiang, Z.~Shao, and J.~Ma, ``Multi-temporal ultra dense
  memory network for video super-resolution,'' \emph{IEEE Transactions on
  Circuits and Systems for Video Technology}, vol.~30, no.~8, pp. 2503--2516,
  2020.

\bibitem{thompson2018freezing}
B.~Thompson, H.~Khayrallah, A.~Anastasopoulos, A.~McCarthy, K.~Duh, R.~Marvin,
  P.~McNamee, J.~Gwinnup, T.~Anderson, and P.~Koehn, ``Freezing subnetworks to
  analyze domain adaptation in neural machine translation,'' in
  \emph{Proceedings of the Third Conference on Machine Translation: Research
  Papers}, Oct. 2018, pp. 124--132.

\bibitem{liu2022video}
H.~Liu, Z.~Ruan, P.~Zhao, C.~Dong, F.~Shang, Y.~Liu, L.~Yang, and R.~Timofte,
  ``Video super-resolution based on deep learning: a comprehensive survey,''
  \emph{Artificial Intelligence Review}, vol.~55, no.~8, pp. 5981--6035, 2022.

\bibitem{girshick2015fast}
R.~Girshick, ``Fast r-cnn,'' in \emph{Proceedings of the IEEE international
  conference on computer vision}, 2015, pp. 1440--1448.

\bibitem{adam}
D.~P. Kingma and J.~Ba, ``Adam: A method for stochastic optimization,''
  \emph{arXiv preprint arXiv:1412.6980}, 2014.

\bibitem{NEURIPS2019_9015}
\BIBentryALTinterwordspacing
A.~Paszke, S.~Gross, F.~Massa, A.~Lerer, J.~Bradbury, G.~Chanan, T.~Killeen,
  Z.~Lin, N.~Gimelshein, L.~Antiga, A.~Desmaison, A.~Kopf, E.~Yang, Z.~DeVito,
  M.~Raison, A.~Tejani, S.~Chilamkurthy, B.~Steiner, L.~Fang, J.~Bai, and
  S.~Chintala, ``Pytorch: An imperative style, high-performance deep learning
  library,'' in \emph{Advances in Neural Information Processing Systems 32},
  H.~Wallach, H.~Larochelle, A.~Beygelzimer, F.~d\textquotesingle
  Alch\'{e}-Buc, E.~Fox, and R.~Garnett, Eds.\hskip 1em plus 0.5em minus
  0.4em\relax Curran Associates, Inc., 2019, pp. 8024--8035. [Online].
  Available:
  \url{http://papers.neurips.cc/paper/9015-pytorch-an-imperative-style-high-performance-deep-learning-library.pdf}
\BIBentrySTDinterwordspacing

\bibitem{sun2016ahg8}
Y.~Sun, A.~Lu, and L.~Yu, ``Ahg8: Ws-psnr for 360 video objective quality
  evaluation,'' in \emph{Joint Video Exploration Team of ITU-T SG16 WP3 and
  ISO/IEC JTC1/SC29/WG11, JVET-D0040, 4th Meeting}, 2016.

\bibitem{lin2017tell}
Y.-C. Lin, Y.-J. Chang, H.-N. Hu, H.-T. Cheng, C.-W. Huang, and M.~Sun, ``Tell
  me where to look: Investigating ways for assisting focus in 360 video,'' in
  \emph{Proceedings of the 2017 CHI Conference on Human Factors in Computing
  Systems}, 2017, pp. 2535--2545.

\bibitem{deng2021lau}
X.~Deng, H.~Wang, M.~Xu, Y.~Guo, Y.~Song, and L.~Yang, ``Lau-net: Latitude
  adaptive upscaling network for omnidirectional image super-resolution,'' in
  \emph{Proceedings of the IEEE/CVF Conference on Computer Vision and Pattern
  Recognition}, 2021, pp. 9189--9198.

\bibitem{ozcinar2019super}
C.~Ozcinar, A.~Rana, and A.~Smolic, ``Super-resolution of omnidirectional
  images using adversarial learning,'' in \emph{2019 IEEE 21st International
  Workshop on Multimedia Signal Processing (MMSP)}.\hskip 1em plus 0.5em minus
  0.4em\relax IEEE, 2019, pp. 1--6.

\bibitem{kappeler2016video}
A.~Kappeler, S.~Yoo, Q.~Dai, and A.~K. Katsaggelos, ``Video super-resolution
  with convolutional neural networks,'' \emph{IEEE transactions on
  computational imaging}, vol.~2, no.~2, pp. 109--122, 2016.

\bibitem{caballero2017real}
J.~Caballero, C.~Ledig, A.~Aitken, A.~Acosta, J.~Totz, Z.~Wang, and W.~Shi,
  ``Real-time video super-resolution with spatio-temporal networks and motion
  compensation,'' in \emph{Proceedings of the IEEE Conference on Computer
  Vision and Pattern Recognition}, 2017, pp. 4778--4787.

\bibitem{FRVSR}
M.~S. Sajjadi, R.~Vemulapalli, and M.~Brown, ``Frame-recurrent video
  super-resolution,'' in \emph{Proceedings of the IEEE Conference on Computer
  Vision and Pattern Recognition}, 2018, pp. 6626--6634.

\bibitem{horizontal_cyc}
M.~Startsev and M.~Dorr, ``360-aware saliency estimation with conventional
  image saliency predictors,'' \emph{Signal Processing: Image Communication},
  vol.~69, pp. 43--52, 2018.

\bibitem{ai2022deep}
H.~Ai, Z.~Cao, J.~Zhu, H.~Bai, Y.~Chen, and L.~Wang, ``Deep learning for
  omnidirectional vision: A survey and new perspectives,'' \emph{arXiv preprint
  arXiv:2205.10468}, 2022.

\end{thebibliography}
\begin{IEEEbiography}[{\includegraphics[width=1in,height=1.25in,clip,keepaspectratio]{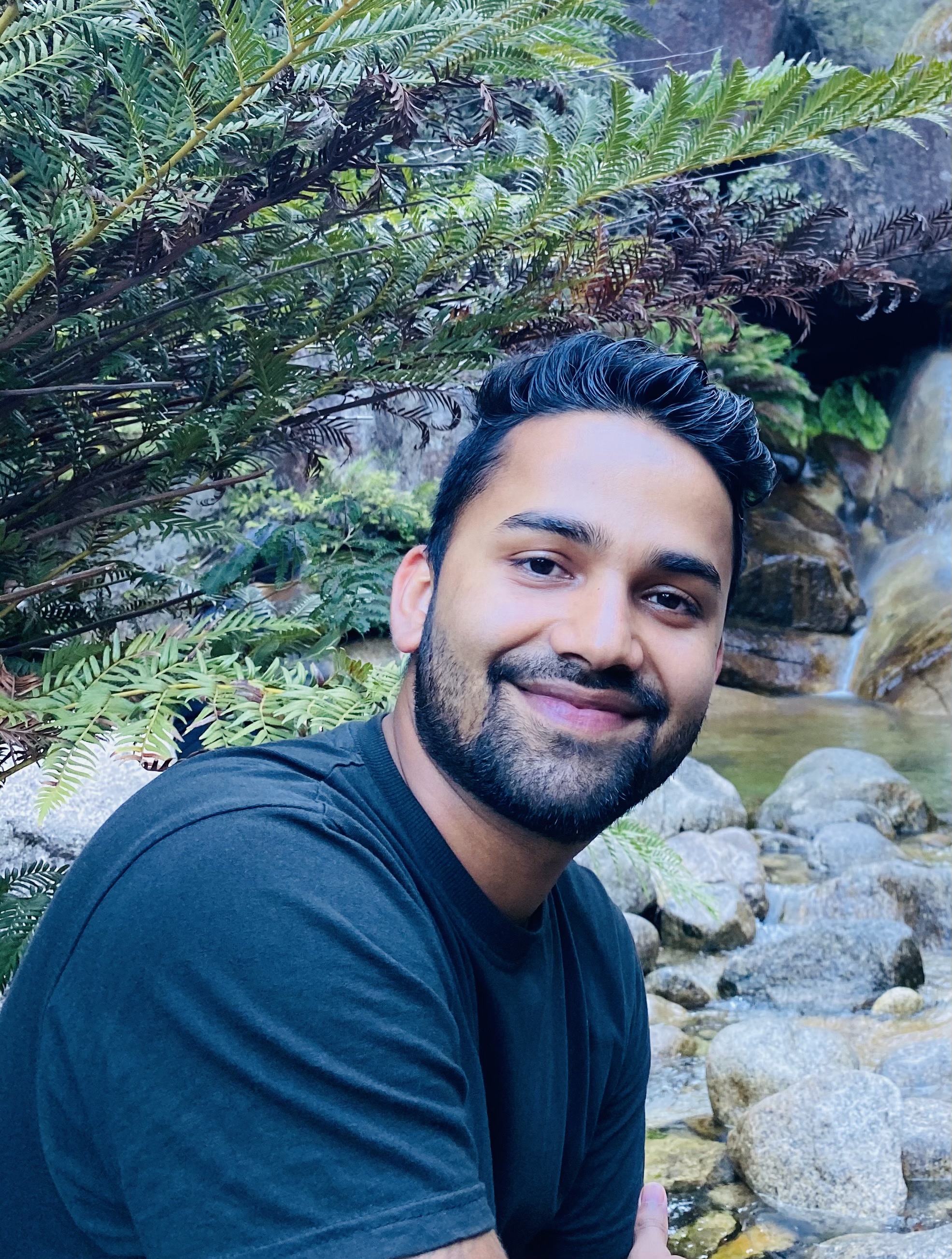}}]{Arbind Agrahari Baniya} is a current PhD candidate at the Centre for Software, Systems and Society (CSSS), School of Information Technology, Deakin University. He is also a Visiting PhD student at Agriculture Victoria Research, Department of Energy, Environment and Climate Action (DECCA) in Victoria, Australia. Prior to this, Arbind completed his Master of Information Technology (Professional) degree from Deakin University in 2019 and worked as Research Fellow at the Center for Regional and Rural Futures (CeRRF), Deakin University. He has also been working as a sessional academic at the School of IT, Deakin University, since 2019. Arbind completed his Bachelor of Technology (Computer Science and Engineering) degree from Jawaharlal Nehru Technological University Kakinada (JNTUK), India, in 2017. He is a recipient of merit-based fully-funded scholarships for both undergraduate and PhD studies. He has also secured funding to undertake a horticulture supply-chain traceability research project for DECCA during his PhD. His research interests include Computer Vision, Machine Learning, Data Science, and, more broadly, Data-Centric Smart Solutions.
\end{IEEEbiography}
\vspace{-3em}
\begin{IEEEbiography}[{\includegraphics[width=1in,height=1.25in,clip,keepaspectratio]{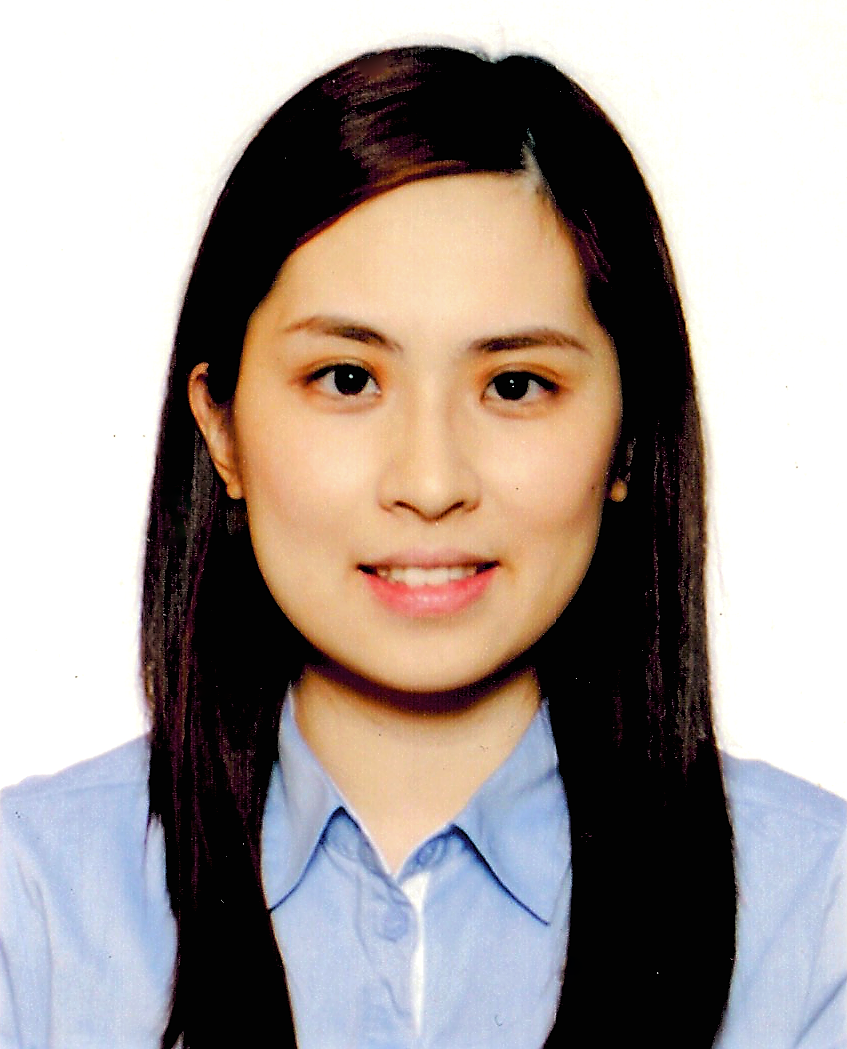}}]{Tsz-Kwan Lee} received her PhD degree from the Department of Electronic and Information Engineering of The Hong Kong Polytechnic University in 2017. During her PhD studies, she was awarded Li Po Chun Charitable Trust Fund Scholarship. Dr Lee was a Postdoctoral Fellow at the Center for Signal Processing at The Hong Kong Polytechnic University in 2017-2018. She was also an invited speaker at the IEEE CAS Guangzhou Workshop 2017, South China University of Technology and the workshop jointly organised by the IEEE CASS Shikoku chapter and IEICE Shikoku chapter, Kagawa University, Takamatsu, Japan. In 2019, Dr Lee joined the School of Computing and Information Systems at the University of Melbourne, Australia. Dr Lee is currently a Lecturer in Computer Science at the School of Information Technology, Deakin University, Australia. In 2022-2023, Dr Lee served as a guest editor for MDPI Applied Sciences Special Issue Challenges and Opportunities in Digital Health. Her research interests include computer vision and applications, multimedia and immersive technologies, machine learning, video quality assessment, 2D/3D video coding, signal processing, and video transcoding.
\end{IEEEbiography}
\vspace{-2em}
\begin{IEEEbiography}[{\includegraphics[width=1in,height=1.25in,clip,keepaspectratio]{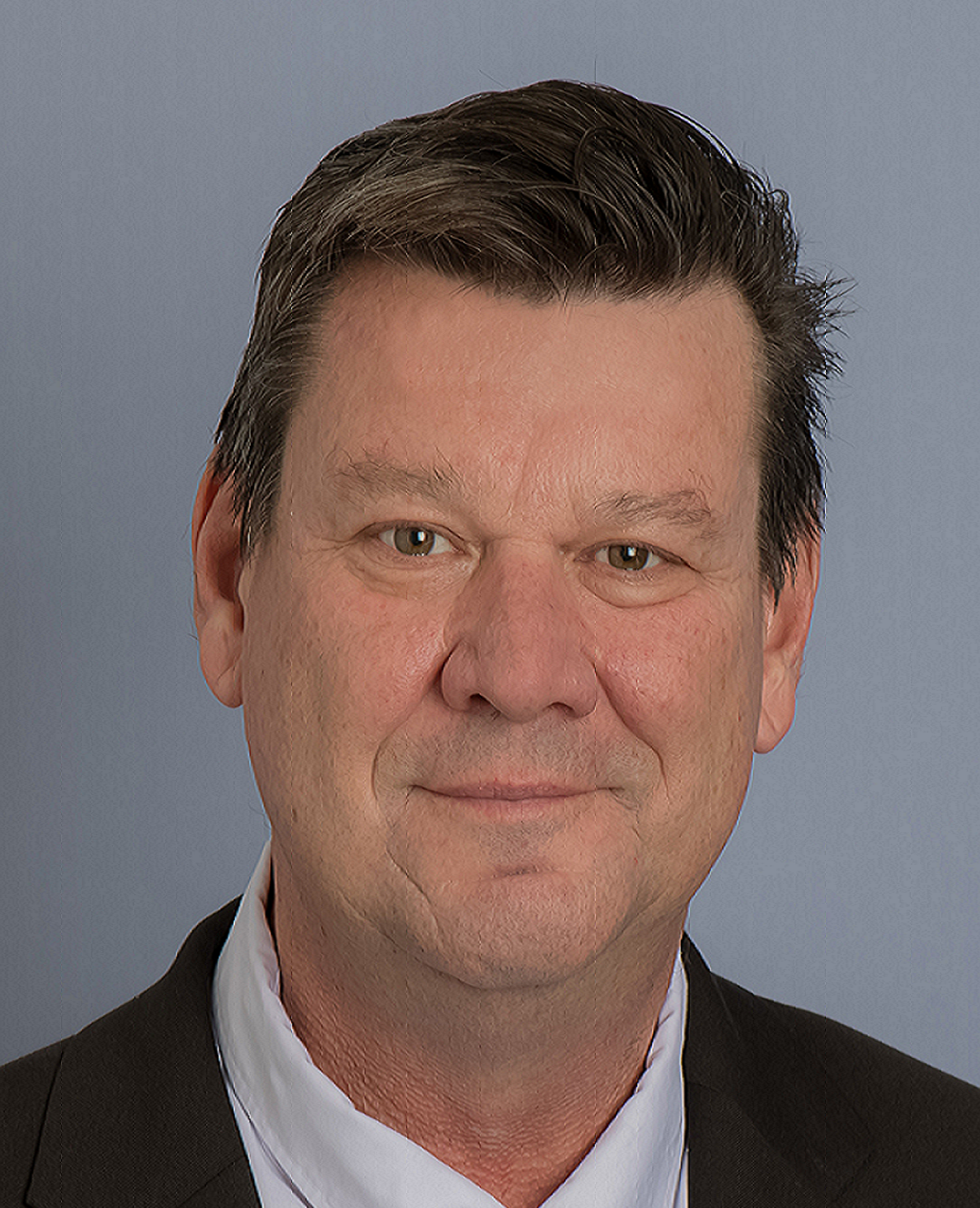}}]{Peter W. Eklund} is Program Lead for Technology Internships at Uniting Vic.Tas, a community services not-for-profit, and Adjunct Professor in Computer Science at Deakin University. An elected fellow of the Australian Computer Society (ACS),  Peter is a tech innovator and computer science scholar. Peter obtained his PhD from Link\"oping University, Sweden (1991),  MPhil from Brighton University, UK (1988) and Bachelor of Mathematics (Honours) from the University of Wollongong, Australia (1985). Peter has enjoyed a 30-year career as a scholar, entrepreneur and consultant in Australia and Europe. In 2012 he won the inaugural Australian Smart Infrastructure Research Award from the Federal Department of Infrastructure, Transport, Regional Development and Local Government.  Peter presently runs several research threads, applications of deep learning methods in NLP, image processing, blockchain and distributed ledgers. Peter is interested in the security of cyber-physical systems that underwrite critical national infrastructure, including supply chains. In 2022 he (and co-authors) won the Andrew P. Sage Award for best IEEE transactions paper. He is also a non-executive director of ZTLment, a Danish  fintech startup company. 
\end{IEEEbiography}
\vspace{-2em}
\begin{IEEEbiography}[{\includegraphics[width=1in,height=1.25in,clip,keepaspectratio]{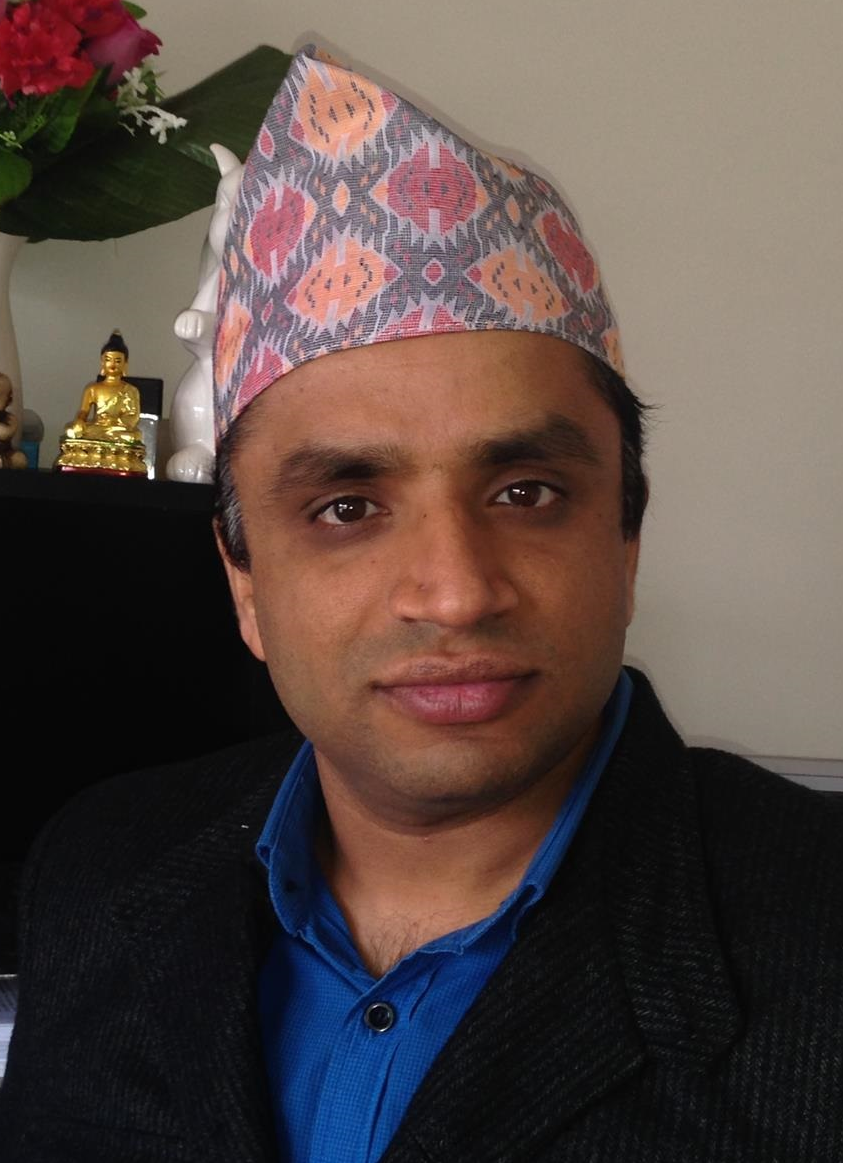}}]{Sunil Aryal}
is a Senior Lecturer in IT at the School of Information Technology, Deakin University, Australia. Prior to joining Deakin University in 2019, he worked as a Lecturer at Federation University Australia and sessional teaching staff at various institutions. Before moving to academia, he worked in the industry as a data engineer and software developer for several years. He received his PhD from Monash University, Australia. His research is in the areas of Data Mining (DM), Machine Learning (ML) and Artificial Intelligence (AI). He is interested in applying ML/AI concepts to solve real-world problems in different domains, particularly in defence, healthcare, cyber security, advanced manufacturing, engineering and IoT. He works on a wide range of areas in ML/AI, from similarity measures, anomaly detection, classification, clustering, text/image analysis, ensemble learning, learning from small subsamples of data, and robust/explainable machine learning to autonomous systems. He has published over 35 scientific papers in top-tier conferences and journals in the field of DM/ML and has been a named investigator on research grants and contracts of over AUD 1.9 million. His research is supported by US Defence (AFOSR and ONR Global), Defence Science and Technology (DST) Group and the Office of National Intelligence (ONI) Australia.
\end{IEEEbiography}

\end{document}